\documentclass[12pt]{amsart}
\usepackage[dvips]{color}
\usepackage{amsmath}
\usepackage{amsxtra}
\usepackage{amscd}
\usepackage{amsthm}
\usepackage{amsfonts}
\usepackage{amssymb}
\usepackage{eucal}
\usepackage{epsfig}
\usepackage{epsf,graphics,mathrsfs,yfonts,eufrak}
\usepackage{braket}
 \usepackage{pdfpages}
 \usepackage{psfrag}
 \usepackage[OT2,T1]{fontenc}
\usepackage{pifont}

\def\'{\char126}
\def\`{\char127}

\def\({\left(}
\def\){\right)}

\newcommand{\be}{{\beta}}
\newcommand{\ka}{{\kappa}}

\newcommand{\ga}{{\gamma}}
\newcommand{\Ga}{{\Gamma}}
\newcommand{\tth}{{\theta}}

\newcommand{\Up}{{\Upsilon}}
\newcommand{\bM}{{\mathbf{M}}}

\newcommand{\Om}{{\Omega}}
\newcommand{\ssh}{{\mathrm{sh}}}

\newcommand{\pa}{{\partial}}
\newcommand{\de}{{\delta}}
\newcommand{\De}{{\Delta}}
\newcommand{\si}{{\sigma}}
\newcommand{\bbe}{{\boldsymbol{\beta}}}
\newcommand{\bga}{{\boldsymbol{\gamma}}}
\newcommand{\bv}{\mathbf{v}}
\newcommand{\bh}{\mathbf{h}}

\newcommand{\vp}{\varphi}

\newcommand{\cb}{\mathbf{c}}
\newcommand{\bb}{\mathbf{b}}

\newcommand{\tb}{\mathbf{t}}

\newcommand{\jb}{\mathbf{j}}

\newcommand{\nn}{\nonumber}

\def\cZ{\mathcal{Z}}

\newcommand{\slth}{\widehat{\mathfrak{sl}}_2}
\newcommand{\res}{{\rm res}}

\newcommand{\Tr}{{\rm Tr}}

\newenvironment{tenumerate}{
  \begin{enumerate}
  
  }{\end{enumerate}}
\newcommand{\bi}{\begin{tenumerate}}
\newcommand{\ei}{\end{tenumerate}}
\newcommand{\isoto}[1][]%
{{\mathop{\buildrel{\sim}\over\longrightarrow}\limits_{#1}}}

\def\[{\left[}
\def\]{\right]}

\newcommand{\al}{\alpha}

\newcommand{\z}{\zeta}
\newcommand{\bs}{\mathbf{s}}

\newcommand{\om}{{\omega}}

\numberwithin{equation}{section}

\newcommand{\cL}{\mathcal{L}}

\newcommand{\bt}{\mathbf{t}}

\def\half{\textstyle{\frac  1 2}}

\newcommand{\bl}{\mathbf{l}}

\newcommand{\bc}{\mathbf{c}}

\def\bi{\mathbf{i}}

\newcommand{\Rho}{\mathrm{P}}

\newcommand{\YN}{\Upsilon_\mathrm{NS}}
\newcommand{\YR}{\Upsilon_\mathrm{R}}

\usepackage[dvipsnames]{xcolor}

\begin{document}
\begin{title}{One point functions of fermionic operators in the Super Sine Gordon model}
\end{title}
\date{\today}
\author{C.~Babenko and  F.~Smirnov}
\address{CB, FS\footnote{Membre du CNRS}: 
 Sorbonne Universite, UPMC Univ Paris 06\\ CNRS, UMR 7589, LPTHE\\F-75005, Paris, France}\email{cbabenko@lpthe.jussieu.fr,smirnov@lpthe.jussieu.fr}

\begin{abstract}
We describe the integrable structure of the space of local operators for 
the supersymmetric sine-Gordon model. Namely,  we conjecture that this space 
is created by acting on the primary fields by  fermions and a Kac-Moody current. 
We proceed with the computation of the one-point functions. In the UV limit they are
shown to agree with the alternative results obtained by solving the reflection relations. 
\end{abstract}

\maketitle
\section{Introduction}

The importance of the one-point functions for the computation of correlation functions
in the framework of the Perturbed Conformal Field Theory (PCFT) is well-known \cite{alzam}.
For the sine-Gordon model at finite temperature the one-point functions were
computed in \cite{HGSV} using the fermionic basis
of the space of local operators. This basis was found first on the lattice
for the (inhomogeneous) six-vertex model \cite{HGSII}. Since  the expectation 
values in the fermionic basis are rather simple the scaling limit is not very difficult to 
consider. One of the main achievements is the exact relation between the local
operators in the fermionic basis and their counterparts in the UV Conformal
Field Theory (CFT).

An alternative approach to the one-point functions uses the reflection relations \cite{FFLZZ,FLZZ1,FLZZ2} which
are based on two reflections (Heisenberg and Virasoro). We do not go into the details which, in addition to
the original papers mentioned above, are discussed in \cite{NeSm}. This way of doing includes certain subtleties 
with the analytical continuation with respect to the coupling constant. However, if the final goal is restricted to
finding a basis in the CFT, invariant under the two reflections, one should not worry because the problem can be
considered as a purely algebraic one. The reflection relations are equivalent to a certain Riemann-Hilbert problem,
and for a long time it was unclear how to solve it. The  synthesis of the two methods, the fermionic basis on the one hand
and the reflection relations on the other, was made in \cite{NeSm}. In this paper it was shown that  the known examples
of the fermionic basis (up to level 8) solve the reflection relations. Moreover, making a qualitative
assumption of  existence of the  fermionic basis one can use the reflection relations in order to compute the fermionic basis quantitatively. 

It is interesting to apply a similar procedure to other integrable models. For the models related to higher ranks
the problem does not look very realistic for the moment. However, the $\slth$ (or rather $U_q(\slth)$) 
symmetric case allows
a highly nontrivial extension to the Fateev model, 
symmetric under   the exceptional  algebra  $U_q(\widehat{D}(2|1;\al))$  \cite{fateev} .
This model deserves the most profound study. It allows numerous particular cases and restrictions. The simplest of
them is the sine-Gordon model (sG) and the next in complexity is the supersymmetric sine-Gordon model (ssG).
The latter is the subject of the present paper. 

Similarly to the sG case we begin the  study of the ssG model by considering its lattice regularization which is the
inhomogeneous 19-vertex model introduced by Fateev and Zamolodchikov \cite{FZ19}
in other words the model based on  the spin-1 evaluation representations of $U_q(\slth)$. 
By the method close to that of the fermionic basis this model was considered in \cite{JMSspin1}
(this paper relies on the previous research \cite{KNS}). Namely, 
it was shown that the
space of (quasi)-local operators allows a basis created by fermions and a Kac-Moody (KM) current
on level one. It is easy to guess that for the integrable lattice models related to higher representations of spin $s$
of 
$U_q(\slth)$ the space of (quasi)-local operators is generated by currents with all half-integer spins up to $s$.
In the scaling limit these models produce the parafermionic sine-Gordon models. If we learn how to 
treat them in their totality it will bring us very close to the general case of the Fateev model.

As has been said, in the present paper we consider the ssG model starting with the 19 vertex lattice model. 
We explain the basis of (quasi)-local operators created by fermions and a KM current at the lattice level. 
Then we proceed to the scaling limit. This provides the basis of local operators for the ssG model
created by fermions and a KM current. 
Our consideration relies heavily on the numerical study of scaling equations for the function 
$\Omega(\theta,\theta')$.
The equation for this function is not rigorously derived, so, it is considered as a conjecture and
should be checked against  alternative data. 
Using the function $\Omega(\theta,\theta')$ it is straightforward to 
compute the one-point functions on the cylinder of radius $R$ (at finite temperature)
for the purely fermionic part of the basis. We restrict our attention to these operators
leaving the KM contributions for future study. 
We consider the UV limit $R\to 0$ in order to find agreement with the corresponding CFT. 

The UV limit is studied using the numerical data and interpolating with respect to the coupling constant,
the quasi-momentum and the parameter of the primary field. There is a difference with the sG case for which
this kind of data allowed to obtain exact relation to the Virasoro descendants up to the level 6. 
Then, an important check of the entire procedure consisted in verifying that the results satisfy the reflection
relations.  
In the ssG case
only level 2 is available by these means. This case agrees with the reflection relations but we would like
to proceed at least a little further. We reverse the procedure following \cite{NeSm}, namely, assuming that there are
local operators created by fermions which transform simply under the reflection, and compute the 
elements of the fermionic basis up to the level 6. 
Needless to say that the reflection relations are considered as a conjecture which is hard to justify rigorously. 

Finally, it is possible to compare with the results obtained by the interpolation
of numerical data finding a perfect agreement. This is the main result of the present paper:
two kind of formulae whose derivations are based on completely different conjectures agree. 

The paper is organized as follows. {In Section \ref{sec:ssG} we introduce the ssG model and recall some previous results, in Section
\ref{sec:Lattice} we describe the 19-vertex model, and introduce the spin 1 fermions and the KM currents.
The function $\Om$ is defined on the lattice and its normalisation is checked. 
In Section \ref{sec:Scaling} the scaling equation for $\Om$ is given, and the first one point functions of fermionic operators obtained by numerical interpolation
are presented. Finally, in Section \ref{sec:Reflection} we describe the alternative approach to the one point functions that uses the reflection relations,
and verify the results obtained in Section \ref{sec:Scaling}. }

\section{Supersymmetric sine-Gordon model}
\label{sec:ssG}

The $N=1 $  supersymmetric sine-Gordon model (ssG) is described by the (Euclidean)  action 
\begin{align}
\mathcal{A}=
\int\Bigl[\Bigl( \frac 1 {4\pi}\partial_z\varphi\partial_{\bar z} \varphi+
\frac 1 {2\pi}\left(\psi\partial_{\bar z}\psi+\bar{\psi}\partial_z\bar{\psi}\right)&-2\mu \bar{\psi}\psi \cos\Bigl(\frac{\beta}{\sqrt{2}}\varphi\Bigr)\Bigr]d^2z\,.\label{action}
\end{align}
Here we take the CFT 
which include massless free boson and Majorana fermion and perturb it by
the operator of dimension  $\Delta=\frac 1 2(1+\beta^2)$. We consider 
the domain $0<\beta<1$ where the perturbation is relevant. 
An additional term needed for the supersymmetry in the classical case is omitted
for known reasons  \cite{BD,lm} . 

The subject of the present paper are the one-point functions,
this corresponds to the  geometry of the cylinder (that we take of radius $R$) with a local insertion.
Correspondingly we consider three types of contours: the contour $c$ encircling the
local insertion and two contours $C_\mp$ which go around the cylinder
to the right and to the left of the insertion. We shall use the notation $C$ talking
about
any of $C_\mp$.
\vskip .5cm
\centerline{\includegraphics{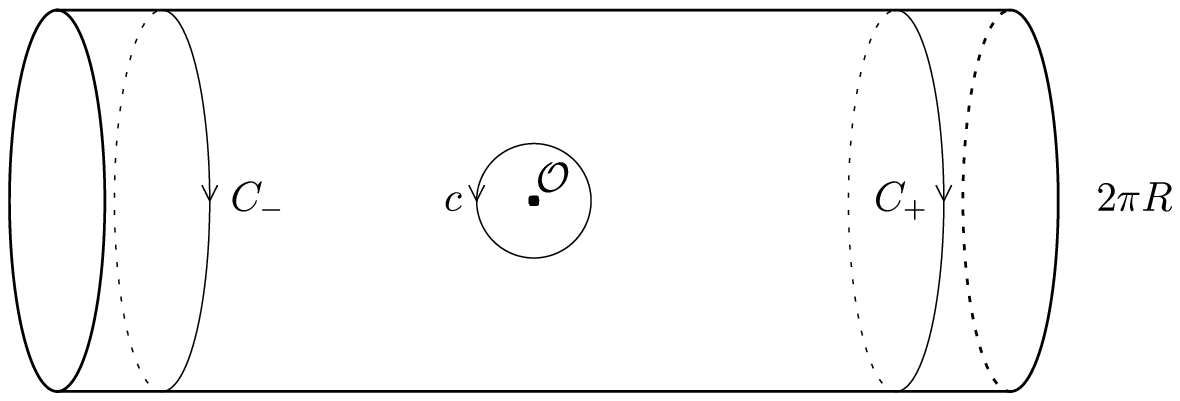}}
\vskip .2cm
\noindent{\it Fig. 1. Cylinder with insertion.}
\vskip .2cm
\noindent
The cylinder is infinite, its generatrix is called Space direction, its directrix is 
called Matsubara direction.
In the present context by the Matsubara transfer-matrix we understand an operator acting from the Matsubara
Hilbert space to itself which is graphically represented as a slice of our cylinder of small Space length $\epsilon$. 
Since the cylinder is infinite, 
both transfer-matrices to the left and
to the right of the insertion are replaced by the  one-dimensional
projectors on the same eigenvector with maximal eigenvalue. 
Since the potential
is invariant under  $\varphi\to\varphi+2\sqrt{2}/\beta$  we can introduce additional
parameter $P$ which is the Floquet index of the Matsubara wave-function. 
The one-point function (partition function with insertion) is denoted by
\begin{align}
\langle \mathcal{O}(0)\rangle_{P,R}\,.\label{OP}
\end{align}

The ssG model can be formally considered as the perturbation of the conformal
complex super-Liouville 
model 
\begin{align}
\mathcal{A}_\mathrm{L}=
\int\Bigl[\Bigl( \frac 1 {4\pi}\partial_z\varphi\partial_{\bar z} \varphi+
\frac 1 {2\pi}\left(\psi\partial_{\bar z}\psi+\bar{\psi}\partial_z\bar{\psi}\right)&-\mu \bar{\psi}\psi e^{-i\frac{\beta}{\sqrt{2}}\varphi}\Bigr)\Bigr]d^2z\,,\label{action1}
\end{align}
by the relevant operator
$$W=\mu \bar{\psi}\psi e^{i\frac{\beta}{\sqrt{2}}\varphi}\,,$$
whose scaling dimension is $\Delta=\beta^2$.

Let us concentrate on the conformal model.
The central charge of the complex super-Liouville 
model  is
$$c={\textstyle \frac 3 2 }\ \hat{c}\,,\quad  
\hat{c}=1-2\bigl({\beta}^{-1}-{\beta}
\bigr)^2\,.$$
In this paper we consider only the NS sector. In the conformal case we can easily change the scale to have $R=1$. 
According to the  usual argument   the operator $\mathcal{O}$ with scaling dimensions ($\Delta_\mathcal{O},
\overline{\Delta}_\mathcal{O}$) in generic position has an uniquely defined counterpart in perturbed theory. We do not
distinguish the two notationally, the UV limit is
\begin{align}
\lim_{r\to 0}r^{\Delta_\mathcal{O}+\overline{\Delta}_\mathcal{O}}\langle \mathcal{O}(0)\rangle_{P,R}=\langle \mathcal{O}(0)\rangle_{P}\,,\label{OP=3P}
\end{align}
{where $r\sim R$ is a dimensionless quantity proportional to $R$, see \eqref{defR} for details.}
By a change of variables from the cylinder to the sphere the CFT one-point function $\langle \mathcal{O}(0)\rangle_{P}$ is 
mapped to the three point function for the image of the operator $\mathcal{O}$ (for descendants this image can
have rather complicated expression in terms of $\mathcal{O}$) and two primary fields with
dimensions
$$ \delta_P= P^2+\frac {\hat{c}-1}{16}\,.$$

The superconformal algebra is generated by the operators $T(z)$, $S(z)$ 
with the OPE's 
\begin{align}
&T(z)T(w)=-\frac{c}{12}\chi'''(z-w)-2\chi'(z-w)T(w)+\chi(z-w)T'(w)+O(1)\,,\label{OPE}\\
&T(z)S(w)=-\frac 3 2 \chi'(z-w)S(w)+\chi(z-w)S'(w)+O(1)\,,\nn\\
&S(z)S(w)= {-} 2\xi(z-w)T(w){-} \frac c  3 \xi''(z-w)+O(1)\,,\nn
\end{align}
where the functions $\chi(z)$, $\xi(z)$ are chosen
to be compatible with our geometry and with the
NS (anti)-periodicity 
coditions:
$$\chi(z)=\frac 1 2\coth\(\frac z {2}\)\,,\quad 
\xi(z)=\frac 1{ 2 \sinh\(\frac z {2}\)}\,.
$$

There are two kinds of the primary fields parametrised by  $\al\in\mathbb{C}$
\begin{align}
V_\al=e^{i\al (\beta^{-1}-\beta)\frac 1{2\sqrt{2}}\varphi}\,,\quad
W_\al= \bar{\psi}\psi e^{i\al (\beta^{-1}-\beta)\frac 1{2\sqrt{2}}\varphi}\,. \label{VWDef}
\end{align}
The  scaling dimension of $V_\al$ is
$$\Delta_\al=\frac 1 8 (\beta^{-1}-\beta)^2\al(\al-2)\,.$$
The scaling dimension of  $W_\al$ equals $\Delta_\al+1/2$.
Later we shall use the OPE's:
\begin{align}
&T(z)V_\al(w)=-\Delta_\al \chi'(z-w)V_{\al}(w)+\chi(z-w)V'_{\al}(w)+O(1)\,,\label{OPEprimary}\\
&T(z)W_\al(w)=-(\Delta_\al+\half) \chi'(z-w)V_{\al}(w)+\chi(z-w)V'_{\al}(w)+O(1)\,,
\nn\\
&S(z)V_\al(w)= {-} \xi(z-w)W_\al(w)+O(1)\,,\nn\\
&S(z)W_\al(w)=  2\Delta_\al \xi'(z-w)V_\al(w) {-} \xi(z-w)V'_\al(w)+O(1)\,,\nn 
\end{align}

Consider the chiral component of the energy momentum tensor $T(z)$. The
geometry of the problem makes it natural to consider two superconformal
algebras
which we call local and global
\begin{align}
&\mathbf{l}_n\mathcal{O}(y)=\frac 1 {2\pi i}\oint_c (z-y)^{n+1}T(z)\mathcal{O}(y)\frac{dz}{2 \pi i }\,,\quad\quad
\ \ \mathbf{s}_k\mathcal{O}(y)=\frac 1 {2\pi i}\oint_c (w-y)^{k+\frac 1 2 }S(w)\mathcal{O}(y)\frac{dw}{2 \pi i } \,, \label{ActionTS} \\
&L_n=\frac{1}{2\pi}\oint_C e^{-{n z}}\Bigl(T(z)+\frac{c}{24}\Bigr)\frac{dz}{2\pi i}\,,
\quad
S_k=\frac{1}{2\pi}\oint_C e^{-{\left(k+\frac{1}{2}\right) z} }S(z)\frac{dz}{2\pi i}\,,
\nn
\end{align} 
where $n\in \mathbb{Z}$, $k\in \mathbb{Z}/2$. The operators $\mathbf{l}_n$,  
$\mathbf{s}_k$ act on the local operators inserted at $z=0$, the operators
$L_n$, $S_k$ act on the Matsubara Hilbert space.

The convenient way of finding the  CFT one-point functions consists in using the OPE and the
asymptotical conditions 
\begin{align}
\lim_{\mathrm{Re}(z)\to\pm \infty}T(z)=\delta_P-\frac {\hat{c}}{16}\,,\quad \lim_{\mathrm{Re}(z)\to\pm \infty}S(z)=0\,.\label{asympt}
\end{align}
In Section \ref{subsec:Descendants} we shall give some examples of computations with these formulae.

The super-Liouville model, in addition to the super-conformal symmetry, possesses
the structure of an integrable model, namely, it allows an infinite number
of local integrals of motion with chiral  local densities $T_{2j}(z), \bar T_{2j}(\bar z)$. 
In our geometry there are two facets of the local integrals of motion: they act  either
on the Matsubara Hilbert space or  on the local
operators inserted at the point $z=0$ being respectively
\begin{align}
I_{2j-1}=\oint_{C}T_{2j}(z)\frac{dz}{2 \pi i }\,,\quad
(\mathbf{i}_{2j-1}\mathcal{O})(0)=\oint_{c}T_{2j}(z)\mathcal{O}(0)\frac{dz}{2\pi i }\,,
\end{align}
and similarly for the other chirality.

Let us write explicitly the first two densities:
\begin{align}
T_2(z)=T(z)\,,\quad T_4(z)=T(z)^2-\frac 1 4 S(z)S'(z)\,.\label{densities}
\end{align}
The formula for $T_2(z)$ means simply that the 
light cone component of the energy-momentum tensor is the first integral, the formula for $T_4(z)$ is the most important: it is well-known that higher local integrals of
motion are completely defined by requirement of commutativity with with the
density  $T_4(z)$.

Let us return to the perturbed model. 
It has been said that
at least for irrational $\al$
the  local operator $V_\al$ and its super-Virasoro
descendants ($W_\al$ in particular) have uniquely  
defined counterparts in the perturbed theory, which we do not 
distinguish notationally. 
The local integrals of motion
survive the perturbation, and get rise to an infinite series of pairs of operators
 $(T_{2j}(z,\bar z),\Theta _{2j-1}(z,\bar z))$,  $(\bar T_{2j}(z,\bar z),\bar\Theta _{2j-1}(z,\bar z))$ satisfying the continuity equations
 $$\partial _{\bar z}T_{2j}(z,\bar z)=\partial_z\Theta _{2j-1}(z,\bar z)\,,\quad
\partial_z\bar T_{2j}(z,\bar z)= \partial _{\bar z}\bar \Theta _{2j-1}(z,\bar z)\,.$$
Consider  $(T_{2j}(z,\bar z),\Theta _{2j-1}(z,\bar z))$,
the other pair being treated quite similarly. 
The action on the local operators is 
\begin{align}
&I_{2j-1}=\frac 1 {2 \pi i }\oint_{C}T_{2j}(z){dz}+\Theta _{2j-1}(z){d\bar{z}}\,,\nn\\
&(\mathbf{i}_{2j-1}\mathcal{O})(0)=\frac 1 {2 \pi i }\oint_{c}T_{2j}(z)\mathcal{O}(0){dz}+\Theta _{2j-1}(z)\mathcal{O}(0){d\bar{z}}\,,\nn
\end{align}
The operators $I_1$, $\bar I_1$  are the light-cone components of the energy-momentum tensor. 

The equivalence
$c=C_--C_+$ implies that the one-point function of a local operators
obtained by the  action of $\mathbf{i}_{2j-1},\bar{\mathbf{i}}_{2j-1}$ vanishes.
So, like in \cite{HGSIV} we work with the quotient space 
$\mathcal{V}_\al^\mathrm{quo}\otimes \overline{\mathcal{V}}_\al^\mathrm{quo}$
obtained from the tensor product of two super conformal Verma modules by factoring
out the descendants of the integrals of motion. The quotient space
$\mathcal{V}_\al^\mathrm{quo}$ will be realised as the one obtained by the action on $V_\al$
of all $\mathbf{s}_{-k}$ and $\mathbf{l}_{-m}$ with even $m$ only.

The particle content of the ssG model consists of solitons and, for 
$\beta^2<1/2$,  their  bound states. There is an exact formula relating
the mass of soliton $M$ to the dimensional coupling constant $\mu$:
\begin{align}
M=\frac{4(1-\beta^2)}{\pi \beta^2}\(\frac{\pi} 2\mu \gamma\Bigl(\frac {1-\beta^2} 2\Bigr)\)^{\frac 1 {1-\beta^2}}\,,\label{mass}
\end{align}
here and later we use the traditional 
notation
$$\gamma(x)=\frac{\Gamma(x)}{\Gamma(1-x)}\,.$$

The free energy of the model is defined by
the maximal  eigenvalue of the Matsbara transfer-matrix. This eigenvalue is found via
the Suzuki equations \cite{suzuki,hrs,BaSm}
\begin{align}
&\log y(\theta)=\int\limits_{-\infty}^{\infty}2\mathrm{Re}\Bigl[L(\theta-\theta'+\pi i \gamma)\log B(\theta'-\pi i \gamma)]
d\theta'\,,\label{eqyfinal}   \\
&\log b(\theta-\pi i\gamma)=-2\pi MR\cosh(\theta-\pi i \gamma)
-\frac{\pi i \sqrt{2}}{\beta}P+\half\log 2\label{eqbfinal}\\&+\int\limits_{-\infty}^{\infty}L(\theta-\theta'+\pi i \gamma)
\log\(\half Y(\theta')\)d\theta' 
\nn\\&+\int\limits_{-\infty}^{\infty}\Bigl[G(\theta-\theta')\log B(\theta'-\pi i \gamma)-G(\theta-\theta'+\pi i (1-2\gamma))\log\overline{B(\theta'-\pi i \gamma)}\Bigr]d\theta'\,.\nn
\end{align}
with kernels being
\begin{align}
&L(\theta)=\frac {1}{2\pi \cosh\theta}\,,\nn\\
&
G(\theta)=\frac 1 {4\pi}\int\limits _{-\infty}^{\infty}\frac 
{\sinh\(\frac{3\beta^2-1}{2(1-\beta^2)}{\pi k}\)}
{\sinh\(\frac{\beta^2}{1-\beta^2}{\pi k}\)\cosh\(\frac{1}{2}\pi k\)}
e^{ik \theta}dk\,.\nn
\end{align}
It is convenient to parametrise $R$ by  the dimensionless  $\theta_0$:
\begin{equation}
R=\frac {\beta}{\sqrt{2}}\(\frac{\pi} 2\mu \gamma\Bigl(\frac {1-\beta^2} 2\Bigr)\)^{-\frac 1 {1-\beta^2}}e^{-\theta_0}\,,
\label{defR}
\end{equation}

The eigenvalues of the local integrals of motion $I_{2k-1}$
are denoted by  $i_{2k-1}$,  they are found from the asymptotics at $\theta\to\infty$
\begin{align}
\log y(\theta)\simeq \sum\limits_{k=1}^{\infty}C_{2k-1}i_{2k-1}(\theta_0)e^{-(2k-1)(\theta-\theta_0)}\,,
\end{align}
and similarly for the asymptotics $\theta\to -\infty$ related to another set of
integrals $\bar I_{2k-1}$. 
In the  limit $\theta_0\to \infty$ the eigenvalue $i_{2k-1}(\theta_0)$ goes to
its CFT value  $i_{2k-1}$, the constants are chosen
as
\begin{align}
C_m=-\sqrt{\frac\pi 2}\ \frac{ \Gamma\(\frac m 2 \)\Gamma\(\frac 1 {1-\beta^2}m\)}{ m!\(\frac{m+1}2\)!\ \beta \ \Gamma\(\frac {\beta^2}{1-\beta^{2}}m\)}\,.
\label{defC}
\end{align}
in order to allow the conventional normalisation of the integrals of motion 
$$i_{2k-1}=P^{2k}+\cdots\,.$$
Exact formulae for $i_1,i_3,i_5$ can be found in \cite{BaSm}.
Notice that contrary to the sG case all the $\Gamma$-functions collapse in the 
formula \eqref{defC} and many similar formulae which we shall have later.

\section{Expectation values in lattice model}
\label{sec:Lattice}
\subsection{General structure}
In the lattice case we historically use for the coupling constant
$$\nu=\frac {1-\beta^2} 2\,.$$

The paper \cite{JMSspin1} considers an (inhomogeneous) 19 vertex 
Fateev-Zamolodchikov model on a cylinder or equivalently 
arbitrary generalised
Gibbs ensemble for the 
(inhomogeneous) spin-1 integrable spin chain.  
In what follows we closely follow the notations of \cite{HGSIII,JMSspin1} with one 
exception: we switch from the multiplicative spectral parameter to the additive one. 
Let us present  some basic formulae. 
As usual we combine the 19 vertices of the model into the $L$-operator
\begin{align}
\cL(\theta)=\left(\begin{array}{ccc|ccc|ccc}
a(\theta)&0&0&0&0&0&0&0&0\\ 
0&b(\theta)&0&c(\theta)&0&0&0&0&0\\
0&0&f(\theta)&0&d(\theta)&0&h(\theta)&0&0\\
\hline
0&c(\theta)&0&b(\theta)&0&0&0&0&0\\
0&0&d(\theta)&0&e(\theta)&0&d(\theta)&0&0\\
0&0&0&0&0&b(\theta)&0&c(\theta)&0\\
\hline
0&0&h(\tth)&0&d(\theta)&0&f(\theta)&0&0\\
0&0&0&0&0&c(\theta)&0&b(\theta)&0\\
0&0&0&0&0&0&0&0&a(\theta)
\end{array}\right)\nn\,,
\end{align}
where
\begin{align}
&a(\theta)=\sinh{\nu} (\theta+{\textstyle\frac {\pi i} 2})\sinh{\nu} \theta\,,\ \ b(\theta)=\sinh{\nu\pi} (\theta-{\textstyle\frac {\pi i} 2})\sinh{\nu} \theta\,,\ \ c(\theta)=\sinh{\nu\pi i }\sinh{\nu} \theta\,,\nn\\
&d(\theta)=\sinh{\nu} (\theta-{\textstyle\frac {\pi i} 2})\sinh{\nu\pi i }\,,\ \ 
f(\theta)=\sinh{\nu} (\theta-{\textstyle\frac {\pi i} 2})\sinh{\nu} (\theta-\pi i )\,,\nn\\
&e(\theta)=\cosh{\nu} (\theta+{\textstyle\frac {\pi i} 2})\cosh{\nu} (\theta-\pi i )-\cosh{\textstyle\frac {\nu\pi i} 2}\,,\ \ h(\theta)=
\sinh {\textstyle\frac {\nu\pi i} 2}\sinh\nu\pi i\,.\nn
\end{align}

We introduce an infinite Space chain of length $N$  and 
the Matsubara chain of length $L$.
Introduce the rectangular monodromy matrix
\begin{align}
&T_{\mathbf{S},\mathbf{M}}=\ \ \ \raisebox{.8cm}{$\curvearrowright $} \hskip -1.05cm\prod\limits_{j=-N/2+1}^{N/2}T_{j,\mathbf{M}}\,,\qquad
T_{j,\mathbf{M}}=\ \raisebox{.7cm}{$\curvearrowleft $} \hskip -.6cm\prod\limits _{m=1}^{L}\cL_{j,m}\,,\nn
\end{align}
where both Space and Matsubara chains can be inhomogeneous,
\begin{align}
\cL_{j,m}=\cL_{j,m}(\xi_j-\tau_m)\,,\label{xxx}
\end{align}
$\xi_j$, ($ \tau_m$) are Space (Matsubara) inhomogenieties. The indices $j,m$ in the right hand side have
double meaning: they count inhomogenieties and the copies in the tensor product. These notations are standard. 
Eventually we take the limit $N\to\infty$, the Space inhomogenieties are suppose to follow 
some regular pattern in the limit. 

Introduce the operators
$$H(j)=\sum\limits _{k=-N/2+1}^jH_j\,,\quad H=H(N/2)\,,$$
with $H_j$ being the Cartan generator acting on $j$-th Space site. 
Consider the ``primary field" $q^{\al H(0)}$, and an operator $\mathcal{O}$ acting 
nontrivially on a finite number of Space sites. 
The operators $q^{\al H(0)}\mathcal{O}$ are called quasi-local. 
The main object of our study is
\begin{align}
\cZ^{\kappa}_{L}\left\{q^{\al H(0)}\mathcal{O}\right\}=\lim_{N\to\infty}
\frac{\Tr_\mathbf{S}\Tr_\mathbf{M}\(T_{\mathbf{S},\mathbf{M}}q^{\kappa H+\al H(0)}\mathcal{O}\)}
{\Tr_\mathbf{S}\Tr_\mathbf{M}\(T_{\mathbf{S},\mathbf{M}}q^{\kappa H+\al H(0)}\)}\,,\label{defZ1}
\end{align}
with $\kappa$ being a parameter. 
Graphically this  is represented on {\it Fig. 2}.
\vskip .5cm
\centerline{\includegraphics{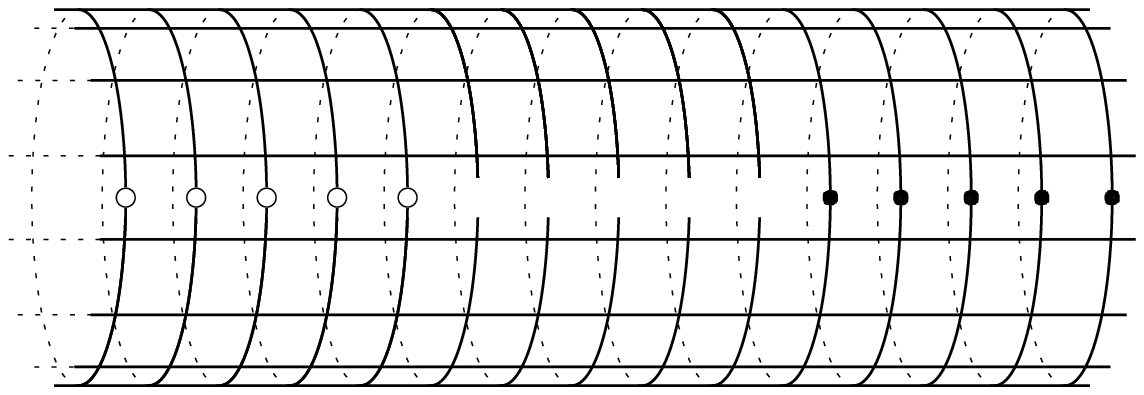}}
\vskip .2cm
\noindent{\it Fig. 2. 19 vertex model on a cylinder with quasi-local insertion.}
\vskip .2cm
\noindent

The main result of \cite{JMSspin1} is that an effective way of computation
goes through the introduction of eight families of creation operators
acting on the space of quasi-local operators. 
These families are
fermions $\bb^*(\theta), \cb^*(\theta)$,  $\tilde{\bb}^*(\theta), \tilde{\cb}^*(\theta)
$\footnote{
These operators were denoted by  $\bar{\bb}^*(\theta), \bar{\cb}^*(\theta)$ in  \cite{JMSspin1}, but we prefer to keep the "bars" for other, more important, use. }, level 1 Kac-Moody currents $\mathbf{j}^+(\theta) $, $\mathbf{j}^-(\theta) $,
$\mathbf{j}^0(\theta) $, and an operator lying in the centre of the entire algebra $\bt^*(\theta)$.  
To be more precise the generating functions of the quasi-local operators are produced by normally 
ordered products of fermions and Kac-Moody currents (the central operator  $\bt^*(\theta)$ does not need normal ordering). This is explained in \cite{JMSspin1,BSspin1}. Since the most significant results of the
present paper concern the quasi-local operators created by fermions only, in which case the normal ordering is not needed, we shall not
go into details of the normal ordering which is standard anyway. 

In the case of homogeneous Space ($\xi_j=0, \forall j$) the creation operators are understood as power
series in $\theta$. 
We shall be interested in the case when the Space inhomogenieties are
staggering: $\xi$ at even sites and $-\xi$ at odd one. In that case every
of above operators give rise to two ``chiral" families defined as power series in $\theta-\xi$, $\theta+\xi$.
All that is absolutely parallel to \cite{HGSV} so we do not go into much details.

The main advantage of our creation operators is that on the descendants which they create acting on
the "primary field", the functional $\cZ^{\kappa}_{L}$ takes simple form. We shall describe
a formal prescription for the computation, detailed explanations being given in \cite{JMSspin1}.
Introduce the creation operators $b^*(\theta)$, $c^*(\theta)$, $t^*(\theta)$, $n(\theta)$
(the first two are fermions, the last two are bosons) which (anti)-commute
among themselves. Prescribe the following values of the functional
$\cZ^{\kappa}_{L}$:
\begin{align}
&\cZ^{\kappa}_{L}\{b^*(\theta^+_1)\cdots b^*(\theta^+_k)c^*(\theta^-_k)\cdots b^*(\theta^-_1)
t^*(\theta^0_1)\cdots t^*(\theta^0_m)n(\sigma_1)\cdots n(\sigma_n)q^{\al H(0)}\}\nn\\
&=\prod_{j=1}^n\frac 1 {\mathcal{N}(\sigma_j)}\prod_{j=1}^m\rho(\theta_j^0)\det\(\omega(\theta_i^+,\theta^-_j)\)_{i,j=1,\cdots k}\,,\nn
\end{align}
where the functions $\mathcal{N}(\theta)$, $\rho(\theta)$, $\omega(\theta,\theta')$ depending on the
Matsubara data will be defined soon. 
The expectation values of the operators created by $\jb^+,\jb^0,\jb^-$, $\bb^*,\cb^*$, $\tilde{\bb}^*,\tilde{\cb}^*$ are computed using the identification
\begin{align}
&\jb^+(\theta)=n(\theta)\ b^*(\theta +{\textstyle \frac{\pi i} 2})b^*(\theta -{\textstyle \frac{\pi i} 2})\,,\label{definitions}\\
&\jb^-(\theta)=n(\theta)\ c^*(\theta -{\textstyle \frac{\pi i} 2})c^*(\theta +{\textstyle \frac{\pi i} 2})\,,\nn\\
&\jb^0(\theta)=n(\theta)\ \bigl(b^*(\theta +{\textstyle \frac{\pi i} 2})c^*(\theta -{\textstyle \frac{\pi i} 2})+c^*(\theta +{\textstyle \frac{\pi i} 2})b^*(\theta -{\textstyle \frac{\pi i} 2})\bigr)\,,\nn\\
&\bb^*(\theta)=n(\theta)\ \bigl(b^*(\theta +{\textstyle \frac{\pi i} 2})t^*(\theta -{\textstyle \frac{\pi i} 2})+b^*(\theta -{\textstyle \frac{\pi i} 2})\bigr)\,,\nn\\
&\cb^*(\theta)=n(\theta)\ \bigl(c^*(\theta +{\textstyle \frac{\pi i} 2})t^*(\theta -{\textstyle \frac{\pi i} 2})+c^*(\theta -{\textstyle \frac{\pi i} 2})\bigr)\,,\nn\\
&\tilde{\bb}^*(\theta)=n(\theta)\ \bigl(b^*(\theta +{\textstyle \frac{\pi i} 2})+t^*(\theta +{\textstyle \frac{\pi i} 2})b^*(\theta -{\textstyle \frac{\pi i} 2})\bigr)\,,\nn\\
&\tilde{\cb}^*(\theta)=n(\theta)\ \bigl(c^*(\theta +{\textstyle \frac{\pi i} 2})+t^*(\theta +{\textstyle \frac{\pi i} 2})c^*(\theta -{\textstyle \frac{\pi i} 2})\bigr)\,.\nn
\end{align}
We had one more operator: $\tb^*(\theta)$, it is similar to $t^*(\theta)$ with $\rho(\theta)$ being
replaced by $\Rho(\theta)$, this function will be given soon. The operator $\tb^*(\theta)$ is in the center, so,
we manipulate it as a $\mathbb{C}$-number. 

\subsection{Basic functions}

The functions $\omega(\theta,\theta')$, $\rho(\theta)$, $\Rho(\theta)$ are defined by the Matsubara
data. The latter consists of the length $L$ chain, with inhomogenieties $\tau_j$, right and left twists 
$\kappa$, $\kappa+\al$, and the eigenvectors with maximal eigenvalues of the right and left transfer-matrices:
$$ T_\bM(\theta|\kappa)=\Tr_j\( T_{j,\bM}(\theta)q^{\kappa H_j}\)\,,\quad
T_\bM(\theta|\kappa+\al)=\Tr_j\( T_{j,\bM}(\theta)q^{(\kappa+\al)H_j}\)\,.$$
Denote the maximal eigenvalues by $T(\theta|\kappa)$, $T(\theta|\kappa+\al)$. Then we are ready to define
the first of our functions:
\begin{align}
\Rho(\theta)=\frac{T(\theta|\kappa+\al)}{T(\theta|\kappa)}\,.\label{defRho}
\end{align}

We shall need the eigenvalues of the two Baxter $Q$-operators \cite{BLZII}
\begin{align}Q^\pm(\theta,\kappa)=e^{\pm\nu\kappa\theta}\prod_{j=1}^m\sinh\nu(\theta-\sigma_j(\kappa))\,,\label{Q}\end{align}
and similarly for $\kappa+\al$. The Bethe roots are denoted by $\sigma_j(\kappa)$. If $\kappa$ is 
not too large the maximal eigenvalue corresponds to $m=L/2$. 
We have the relation between $T$ and $Q^\pm$:
\begin{align}
T(\theta,\kappa)&=a(\theta+\pi i /2)a(\theta-\pi i /2)\frac{Q^\pm(\theta+3\pi i /2,\kappa)}{Q^\pm(\theta-\pi i /2,\kappa)}\nn\\&+a(\theta+\pi i /2)d(\theta-\pi i /2)\frac{Q^\pm(\theta-3\pi i /2,\kappa)Q^\pm(\theta+3\pi i /2,\kappa)}{Q^\pm(\theta-\pi i /2,\kappa)Q^\pm(\theta+\pi i /2,\kappa)}\nn\\&+d(\theta+\pi i /2)d(\theta-\pi i /2)\frac{Q^\pm(\theta-3\pi i /2,\kappa)}{Q^\pm(\theta+\pi i /2,\kappa)}\,,\nn
\end{align}
where
$$a(\theta)=s(\theta-\pi i),\quad d(\theta)=s(\theta+\pi i),\quad s(\theta)=\prod_{j=1}^L\sinh\nu(\theta-\tau_j)\,.$$
We shall need the eigenvalues of the transfer-matrix with the two-dimensional auxiliary space $t(\theta|\kappa)$, for which 
\begin{align}
&t(\theta,\kappa)Q^\pm(\theta,\kappa)=a(\theta)Q^\pm(\theta+\pi i,\kappa)+d(\theta)Q^\pm(\theta-\pi i,\kappa)\,,\nn\\
&T(\theta,\kappa)=t(\theta-\pi i /2,\kappa)t(\theta+\pi i /2,\kappa)-f(\theta)\,,\quad f(\theta)=s(\theta-3\pi i/2)s(\theta+3\pi i/2)\,.\nn
\end{align}
Denote
\begin{align}y(\theta)=\frac{T(\theta|\kappa)}{f(\theta)}\,.\label{defy}\end{align}
We do not explicitly indicate the dependence of $y(\theta)$ on $\kappa$ because
it will be never used for another value of twist.

Now we are ready to define two more  functions
\begin{align}
\rho(\theta)=\frac{t(\theta|\kappa+\al)}{t(\theta|\kappa)}\,,\quad \mathcal{N}(\theta)=\frac{y(\theta)}{1+y(\theta)}\,.\label{defrho}
\end{align}

Recall the definition of the function $\om$ from \cite{HGSIII}. This function splits in two parts: 
\begin{equation}
\om(\tth,\tth') = \om_{\mathrm{hol}}(\tth,\tth') + \om_{\mathrm{sing}}(\tth,\tth')  \,,
\end{equation}
where $\omega_\mathrm{hol}(\theta,\tth')$ as a function of $\theta$ has no other singularities  but simple poles at the zeros of $T(\theta,\kappa)$, 
and $\om_{\mathrm{sing}}$  is its singular part given by  : 
\begin{align}
&\omega_\mathrm{sing}(\theta,\theta')=
\frac 1 {t(\theta|\kappa)t(\theta'|\kappa)}\Bigl(a(\theta)d(\theta')\psi(\theta-\theta'+\pi i,\al)-d(\theta)a(\theta')\psi(\theta-\theta'-\pi i,\al)\Bigr)  \label{omSing}  \\
&+(1+\rho(\tth)\rho(\tth'))\phi(\theta-\theta',\al)-\rho(\tth)\phi(\theta-\theta'+\pi i,\al)-\rho(\tth')\phi(\theta-\theta'-\pi i,\al)\,,\nn
\end{align}
where 
\begin{equation}
 \psi(\theta,\al)=2\nu\frac{e^{\al\nu\theta} }{ e^{2\nu\theta}-1 } 
 \,,
\end{equation}
and $\phi$ is defined as a solution of the difference equation: 
\begin{equation}
\De_{\tth} \phi(\tth,\al)  = \phi(\tth+i\pi,\al)- \phi(\tth-i\pi,\al) = \psi(\tth,\al) \,.
\end{equation}

We shall remind the  normalisation conditions for  the function $\om$.  Start by defining the function $\varphi$:
$$\varphi(\theta)=\prod_{j=1}^L\frac 1{\sinh\nu(\theta-\tau_j-\pi i)\sinh\nu(\theta-\tau_j)\sinh\nu(\theta-\tau_j+\pi i )}\,,$$ 
satisfying
$$d(\theta+\pi i)\varphi(\theta+\pi i)=a(\theta)\varphi(\theta)\,,$$
and the measure
\begin{align}
d\mu^\pm(\theta)=Q^\mp(\theta,\kappa+\al)Q^\pm(\theta,\kappa)\varphi(\theta)d\theta\,.\label{measure}
\end{align}
The poles of 
$\varphi$ come in  triplets 
reflecting the fact that the Matsubara chain consists of
spin-1 representations. 
Let the contour $\Gamma_j$ go around the three points $\tau_j,\tau_j\pm\pi i $.
The normalisation conditions on the  function $\omega(\theta,\eta)$ from \cite{HGSIII} are given by : 
\begin{align}
\int\limits_{\Gamma_j}t(\theta,\kappa)\omega(\theta,\eta)d\mu^+(\theta)=0\,,\label{norm}
\end{align}
The equations \eqref{omSing} , \eqref{norm} define $\omega(\theta,\theta')$ completely. 
Due to the deformed Riemann bilinear identity the following relation is automatic:
$$\int\limits_{\Gamma_j}t(\theta,\kappa)\omega(\eta,\theta)d\mu^-(\theta)=0\,.$$

In order to make the further formulae more readable we shall denote by $\tau$ without index any of
inhomogenieties $\tau_j$. 

For future use we rewrite the normalisation condition as 
\begin{align}
\omega(\tau+\pi i,\eta)
&+Y(\tau)\omega(\tau,\eta)+X(\tau)\omega(\tau-\pi i,\eta)=0\,,\label{norm1}
\end{align}
with
\begin{align}
&X(\tth)=\frac {T(\tth+\pi i /2,\ka+\al)}{a(\tth)d(\tth+\pi i )}\,,
\quad\quad Y(\tth) = \frac{1}{\rho(\tth)}\(1+X(\tth) \)
\,.\nn
\end{align}
Similarly,
\begin{align}
\omega(\eta,\tau+\pi i)
&+Y(\tau)\omega(\eta,\tau)+X(\tau)\omega(\eta,\tau-\pi i)=0\label{norm2}\,,
\end{align}

\subsection{Rewriting normalisation conditions} 
Introduce
\begin{align}
&F^+(\theta,\eta)=\langle \bb^*(\theta)c^*(\eta)\rangle=\frac 1 {\mathcal{N}(\theta)}\Bigl(\omega(\theta+\pi i/2,\eta)\rho(\theta-\pi i/2)+
\omega(\theta-\pi i/2,\eta)
\Bigr)\,,\nn\\
&\tilde{F}^+(\theta,\eta)=\langle \tilde{\bb}^*(\theta)c^*(\eta)\rangle=\frac 1 {\mathcal{N}(\theta)}\Bigl(\omega(\theta+\pi i/2,\eta)+\rho(\theta+\pi i/2)
\omega(\theta-\pi i/2,\eta)\,,
\Bigr)\,,\nn\\
&F^-(\eta,\theta)=\langle b^*(\eta)\bc^*(\theta)\rangle=\frac 1 {\mathcal{N}(\theta)}\Bigl(\omega(\eta,\theta+\pi i/2)\rho(\theta-\pi i/2)+
\omega(\eta,\theta-\pi i/2)
\Bigr)\,,\nn\\
&\tilde{F}^-(\eta,\theta)=\langle b^*(\eta)\tilde{\bc}^*(\theta)\rangle=\frac 1 {\mathcal{N}(\theta)}\Bigl(\omega(\eta,\theta+\pi i/2)+\rho(\theta+\pi i/2)
\omega(\eta,\theta-\pi i/2)
\Bigr)\,.\nn
\end{align}

The functions describe pairings between the fused operators $ \bb^*(\theta)$,  $\cb^*(\theta)$ with
unfused ones $c^*$, $b^*$. Clearly knowledge of these pairings is sufficient to compute any expectation
value containing $ \bb^*(\theta)$,  $\cb^*(\theta)$. So, the analytical properties of $F^+(\theta,\eta)$ {\it etc} characterise in the weak sense the analytical properties of  $ \bb^*(\theta)$,  $\cb^*(\theta)$.

Similarly, in order to understand the analytical properties of $\mathbf{j}^+(\theta),\mathbf{j}^0(\theta),\mathbf{j}^-(\theta)$ we introduce
\begin{align}
G^+(\theta,\eta_1,\eta_2)&=\langle \mathbf{j}^+(\theta)c^*(\eta_2)c^*(\eta_1)\rangle=\frac 1 {\mathcal{N}(\theta)}\left|\begin{matrix}\omega(\theta+\pi i/2,\eta_1)&\omega(\theta+\pi i/2,\eta_2)\\\omega(\theta-\pi i/2,\eta_1)&\omega(\theta-\pi i/2,\eta_2)
\end{matrix}\right|\,,\nn\\
G^-(\eta_1,\eta_2,\theta)&=\langle b^*(\eta_1)b^*(\eta_2)\mathbf{j}^-(\theta)\rangle=\frac 1 {\mathcal{N}(\theta)}\left|\begin{matrix}\omega(\eta_1,\theta+\pi i/2)&\omega(\eta_2,\theta+\pi i/2)\\\omega(\eta_1,\theta-\pi i/2)&\omega(\eta_2,\theta-\pi i/2)
\end{matrix}\right|\,,
\nn\\
G^0(\theta,\eta_1,\eta_2)\ &=\langle\mathbf{j}^0(\theta) b^*(\eta_1)c^*(\eta_2)\rangle\nn\\&=\frac 1 {\mathcal{N}(\theta)}\Bigl((\omega(\theta+\pi i /2,\theta-\pi i /2)-\omega(\theta-\pi i /2,\theta+\pi i /2))
\omega(\eta_1,\eta_2)\nn\\&+\omega(\theta-\pi i/2,\eta_2)\omega(\eta_1,\theta+\pi i/2)-\omega(\theta+\pi i/2,\eta_2)\omega(\eta_1,\theta-\pi i/2)\Bigr)\,,
\nn
\end{align}
where in the last line we imply
$$\omega(\theta+\pi i /2,\theta-\pi i /2)-\omega(\theta-\pi i /2,\theta+\pi i /2)=\lim_{\theta'\to\theta}\Bigl(\omega(\theta+\pi i /2,\theta'-\pi i /2)-\omega(\theta-\pi i /2,\theta'+\pi i /2)\Bigr)\,.$$

We want to rewrite the normalisation conditions in terms of these functions and $\Rho(\theta)$ only.
As before let $\tau$ be any inhomogeniety. Then we claim that
\begin{align}
&F^+(\tau+\pi i /2,\eta)+\Rho(\tau+\pi i /2)\tilde{F}^+(\tau-\pi i /2,\eta)=0\,.\label{eq1}\\
&\tilde{F}^+(\tau+\pi i /2,\eta)+\Rho(\tau+\pi i /2)F^+(\tau-\pi i /2,\eta)=0\,,\nn\\
&F^-(\eta,\tau+\pi i /2)+\Rho(\tau+\pi i /2)\tilde{F}^-(\eta,\tau-\pi i /2)=0\,,\nn\\
&\tilde{F}^-(\eta,\tau+\pi i /2)+\Rho(\tau+\pi i /2)F^-(\eta,\tau-\pi i /2)=0\,,\nn\\
&G^+(\tau+\pi i /2,\eta_1,\eta_2)-\Rho(\tau+\pi i /2)G^+(\tau-\pi i /2,\eta_1,\eta_2)=0\,,\nn\\
&G^-(\tau+\pi i /2,\eta_1,\eta_2)-\Rho(\tau+\pi i /2)G^-(\tau-\pi i /2,\eta_1,\eta_2)=0\,,\nn\\
&G^0(\tau+\pi i /2,\eta_1,\eta_2)-\Rho(\tau+\pi i /2)G^0(\tau-\pi i /2,\eta_1,\eta_2)=0\,.\nn
\end{align}
Let us prove the first of these identities, others are checked similarly.

We begin with some useful identities. 
Using
$$ a(\tau+\pi i)=0\,,\qquad d(\tau-\pi i)=0\,,$$
we find 
\begin{align}
&t(\tau+\pi i,\kappa)=d(\tau+\pi i)\frac{Q^{\pm}(\tau,\kappa)}{Q^{\pm}(\tau+\pi i,\kappa)}
\,,\label{degenerate}\\
&t(\tau-\pi i,\kappa)=a(\tau-\pi i)\frac{Q^{\pm}(\tau,\kappa)}{Q^{\pm}(\tau-\pi i,\kappa)}
\,,\nn\\
&T(\tau+\pi i/2,\kappa)=d(\tau+\pi i)d(\tau)\frac{Q^{\pm}(\tau-\pi i ,\kappa)}{Q^{\pm}(\tau+\pi i,\kappa)}
\,,\nn\\
&T(\tau-\pi i/2,\kappa)=a(\tau-\pi i)a(\tau)\frac{Q^{\pm}(\tau+\pi i ,\kappa)}{Q^{\pm}(\tau-\pi i,\kappa)}
\,.\nn
\end{align}

For \eqref{eq1} we have
\begin{align}
&F^+(\tau+\pi i /2,\eta)+\Rho(\tau+\pi i /2)\tilde{F}^+(\tau-\pi i /2,\eta)\nn\\
&=\frac 1 {\mathcal{N}(\tau+\pi i /2)}\Bigl(
\omega(\tau+\pi i,\eta)\rho(\tau)+\omega(\tau,\eta)\Bigl(1+\frac  {\mathcal{N}(\tau+\pi i /2)}{\mathcal{N}(\tau-\pi i /2)}\Rho(\tau+\pi i /2)
\Bigr)\nn\\&+\omega(\tau-\pi i,\eta)\rho(\tau)\frac  {\mathcal{N}(\tau+\pi i /2)}{\mathcal{N}(\tau-\pi i /2)}\Rho(\tau+\pi i /2)\Bigr)\,.\nn
\end{align}
Using \eqref{degenerate} we compute
\begin{align}
&\frac  {\mathcal{N}(\tau+\pi i /2)}{\mathcal{N}(\tau-\pi i /2)}\Rho(\tau+\pi i /2)
=\frac{d(\tau)}{a(\tau)}\frac {Q^-(\tau-\pi i ,\kappa+\al)}{Q^-(\tau+\pi i ,\kappa+\al)}=X(\tau)\,.\nn
\end{align}
Using the latter identity
we evaluate
\begin{align}
&F^+(\tau+\pi i /2,\eta)+\Rho(\tau+\pi i /2)\tilde{F}^+(\tau-\pi i /2,\eta)\nn\\
&=\frac {\rho(\tau)} {\mathcal{N}(\tau+\pi i /2)}\Bigl(
\omega(\tau+\pi i,\eta)+\omega(\tau,\eta)\frac {Q^-(\tau ,\kappa+\al)T_1(\tau ,\kappa+\al)}{a(\tau)\rho(\tau)Q^-(\tau+\pi i ,\kappa+\al)}\nn\\&+X(\tau)\omega(\tau-\pi i,\eta)\Bigr)\nn\\
&=\frac {\rho(\tau)} {\mathcal{N}(\tau+\pi i /2)}\Bigl(
\omega(\tau+\pi i,\eta)+Y(\tau)\omega(\tau,\eta)+X(\tau)\omega(\tau-\pi i,\eta)\Bigr)=0
\,,\nn
\end{align}
due to \eqref{norm}.

\subsection{The case $\al=0$}

In the case $\al=0$ the left and right eigenstates coincide, hence $\rho(\theta)=1$ and
in the weak there is no difference between $\bb^*,\cb^*$ at one hand and $\tilde\bb^*,\tilde\cb^*$
on the other. So, all the expectation values containing only fermions
are expressed via one function
\begin{align}
&\Omega(\theta,\theta')=\frac 1 {\mathcal{N}(\theta)\mathcal{N}(\theta')}\label{defOmega}\\&\times\Bigl(
\omega(\theta+{\textstyle\frac {\pi i}2},\theta'+{\textstyle\frac {\pi i}2})
+\omega(\theta+{\textstyle\frac {\pi i}2},\theta'-{\textstyle\frac {\pi i}2})
+\omega(\theta-{\textstyle\frac {\pi i}2},\theta'+{\textstyle\frac {\pi i}2})
+\omega(\theta-{\textstyle\frac {\pi i}2},\theta'-{\textstyle\frac {\pi i}2})
\Bigr)\,,\nn
\end{align}
We want to find an independent way of defining this function. 
As explained in \cite{HGSIII} for $\al=0$ there is an important analogy between
the function $\omega(\theta,\theta')$ and the normalised second kind differential on
a hyperelliptic Riemann surface. The normalisation condition \eqref{norm}
is the analogue of the requirement of vanishing of the $a$-periods. 

We set
$$\tau=\tau_j\,.$$

Consider the function 
$$
\tilde\omega(\theta)=\frac{\delta}{\delta\tau}\Bigl\{\log t(\theta)-
\log\Bigl(\frac{s(\theta-\pi i)s(\theta+\pi i )}{s(\theta)}\Bigr)\Bigr\}
\Bigr)\,.
$$ 
Notice that
$$\frac{\delta}{\delta\tau}\log\Bigl(\frac{s(\theta-\pi i)s(\theta+\pi i )}{s(\theta)}\Bigr)=(\delta^+_\theta)^{-1}
\frac{\delta}{\delta\tau}\log\bigl( s(\theta-2\pi i)s(\theta+\pi i )\bigr)\,,$$
where $\delta^+_\theta f(\theta)=f(\theta)+f(\theta-\pi i)$.

We want to show that $\tilde\omega(\theta)$ is a normalised differential.
First we prove that
$$\int_{\Gamma_{k}}t(\theta)\tilde\omega(\theta)d\mu^\pm(\theta)=0\,,\quad k\ne j\,.$$
The case $k=j$ is special, instead of direct computation for this case we consider
$\Gamma_{\pm\infty}=[\pm\Lambda,\pm\Lambda+\pi i/\nu]$ for $|\lambda|>\max(|\tau_k|)$. For $\Gamma_{\pm\infty}$ the computation
is exactly the same as for $\Gamma_k$, $k\ne j$. 

Recall that (in the case $\al = 0$ we have $d\mu^+ = d\mu^- = d\mu$) :
$$d\mu(\theta)=Q^+(\theta)Q^-(\theta)\varphi(\theta)d\theta\,.$$
We have two identities \cite{HGSIII}:
\begin{align}
\int_{\Gamma_{k}}t(\theta)(\delta^+_\theta)^{-1}f(\theta)d\mu(\theta)&=
\int_{\Gamma_{k}}d(\theta)f(\theta)Q^+(\theta -\pi i )Q^-(\theta)\varphi(\theta)d\theta\,,\nn\\
&
=
\int_{\Gamma_{k}}a(\theta)f(\theta +\pi i)Q^+(\theta +\pi i)Q^-(\theta)\varphi(\theta)d\theta\,.\nn
\end{align}
Using these identities we derive
\begin{align}
&\int_{\Gamma_{k}}t(\theta)\tilde\omega(\theta)d\mu(\theta)\nn\\
&=\int_{\Gamma_{k}}\Bigl\{Q^+(\theta)\frac{\delta}{\delta\tau}t(\theta)-Q^+(\theta +\pi i )\frac{\delta}{\delta\tau}a(\theta)-Q^+(\theta -\pi i)\frac{\delta}{\delta\tau}d(\theta)
\Bigr\}Q^-(\theta)\varphi(\theta)d\theta\nn\\&=\int_{\Gamma_{k}}\Bigl\{a(\theta)\frac{\delta}{\delta\tau}Q^+(\theta +\pi i )+d(\theta)\frac{\delta}{\delta\tau}Q^+(\theta -\pi i)-t(\theta)\frac{\delta}{\delta\tau}Q^+(\theta)
\Bigr\}Q^-(\theta)\varphi(\theta)d\theta\nn\\&=
\int_{\Gamma_{k}}\Bigl\{a(\theta)Q^-(\theta)\frac{\delta}{\delta\tau}Q^+(\theta +\pi i )-d(\theta)Q^-(\theta -\pi i)\frac{\delta}{\delta\tau}Q^+(\theta)\Bigr\}\varphi(\theta)d\theta\nn
\\&+\int_{\Gamma_{k}}\Bigl\{d(\theta)Q^-(\theta)\frac{\delta}{\delta\tau}Q^+(\theta -\pi i)-a(\theta)Q^-(\theta +\pi i )\frac{\delta}{\delta\tau}Q^+(\theta)
\Bigr\}\varphi(\theta)d\theta=0\,.\nn
\end{align}
As a normalised differential $\tilde\omega(\theta)$ must be expressible as a linear combination of $\omega(\theta,\eta_j)$ for some 
set $\{\eta_j\}$. The structure of singularities of  $\tilde\omega(\theta)$ suggests that this set is just $\tau,\tau+\pi i$. To be
precise we claim that
\begin{align}
\tilde\omega(\z)=\frac 1 {\mathcal{N}(\tau+\frac{\pi i} 2)}(\omega(\z,\tau)+\omega(\z,\tau+\pi i))\,.\label{o=o+o}
\end{align}
Let us prove this. We have
\begin{align}
\omega(\theta,\tau)+\omega(\theta,\tau+\pi i)=\omega_\mathrm{hol}(\theta,\tau)+\omega_\mathrm{hol}(\theta,\tau+\pi i)+
\omega_\mathrm{sing}(\theta,\tau)+\omega_\mathrm{sing}(\theta,\tau+\pi i)\,,\nn
\end{align}
where $\omega_\mathrm{hol}(\theta,\eta)$as function of $\theta$ has no other singularities but simple poles at
zeros of $t(\theta)$,
\begin{align}\omega_\mathrm{sing}(\theta,\eta)&=\delta^-_\theta\delta^-_\eta\Delta^{-1}_\theta (\nu\coth\nu(\theta-\eta))\nn\\&+
\frac 1 {t(\theta)t(\eta)}\bigl(a(\theta)d(\eta)\nu\coth\nu(\theta-\eta+\pi i)-d(\theta)a(\eta)\nu\coth\nu(\theta-\eta-\pi i)\bigr)\,,
\end{align}
which implies
\begin{align}
&\omega_\mathrm{sing}(\theta,\tau)+\omega_\mathrm{sing}(\theta,\tau+\pi i)=\nu\coth\nu(\theta-\tau-\pi i)-\nu\coth\nu(\theta-\tau)\nn\\&
+\frac{a(\theta)d(\tau)}  {t(\theta)t(\tau)}\nu\coth\nu(\theta-\tau+\pi i)-\frac{d(\theta)a(\tau)}{t(\theta)t(\tau)}\nu\coth\nu(\theta-\tau-\pi i)\nn\\&+\frac{a(\theta)d(\tau+\pi i )}  {t(\theta)t(\tau+\pi i )}\nu\coth\nu(\theta-\tau)\,.\nn\\
&=\frac{a(\theta)d(\tau)}  {t(\theta)t(\tau)}\nu\coth\nu(\theta-\tau+\pi i)+\frac{t(\theta)t(\tau)-d(\theta)a(\tau)}{t(\theta)t(\tau)}\nu\coth\nu(\theta-\tau-\pi i)\nn\\&-\frac{t(\theta)t(\tau+\pi i )-a(\theta)d(\tau+\pi i )}  {t(\theta)t(\tau+\pi i )}\nu\coth\nu(\theta-\tau)\,.\nn
\end{align}
Using this identity one finds
\begin{align}
&\res_{\theta=\tau-\pi i}\bigl(\omega_\mathrm{sing}(\theta,\tau)+\omega_\mathrm{sing}(\theta,\tau+\pi i)\bigr)=
\frac{a(\tau-\pi i)d(\tau)}{t(\tau-\pi i)t(\tau)}=\mathcal{N}(\tau+\frac{\pi i} 2)\,,\nn\\
&\res_{\theta=\tau+\pi i}\bigl(\omega_\mathrm{sing}(\theta,\tau)+\omega_\mathrm{sing}(\theta,\tau+\pi i)\bigr)=
\frac{T_2(\tau+\frac{\pi i } 2 )}{t(\tau+\pi i)t(\tau)}=\mathcal{N}(\tau+\frac{\pi i} 2)\,,\nn\\
&\res_{\theta=\tau}\bigl(\omega_\mathrm{sing}(\theta,\tau)+\omega_\mathrm{sing}(\theta,\tau+\pi i)\bigr)=
-\frac{T_2(\tau+\frac{\pi i } 2)}{t(\tau+\pi i)t(\tau)}=-\mathcal{N}(\tau+\frac{\pi i} 2)\,,\nn
\end{align}
This finishes the proof.

Now we obtain the most important relation of this section : 
\begin{align}
&\frac{\delta}{\delta\tau}\log\(\frac{ T_2(\theta)}{f(\theta)}\)\label{dy=O}\\&=\frac 1 {\mathcal{N}(\theta)}\(\frac{\delta}{\delta\tau}\log T_1(\theta+\pi i/2)+\frac{\delta}{\delta\tau}\log T_1(\theta-\pi i/2)\)-
\(\frac{f(\theta)} {T_2(\theta)}+1\)\frac{\delta}{\delta\tau}\log f(\theta)\nn  \\
&=\frac 1 {\mathcal{N}(\theta)}\(\frac{\delta}{\delta\tau}\log T_1(\theta+\pi i/2)+\frac{\delta}{\delta\tau}\log T_1(\theta-\pi i/2)-\frac{\delta}{\delta\tau}\log f(\theta)\)
\nn\\
&=\frac 1 {\mathcal{N}(\theta)\mathcal{N}(\tau+\frac{\pi i} 2)}
\(\omega(\theta+\frac{\pi i} 2,\tau)+\omega(\theta+\frac{\pi i} 2,\tau+\pi i)+\omega(\theta-\frac{\pi i} 2,\tau)+\omega(\theta-\frac{\pi i} 2,\tau+\pi i)\)\,.\nn\\
&= \Omega(\theta,\tau+\frac{\pi i } 2)\,.  \nn
\end{align}

\section{Scaling limit}
\label{sec:Scaling}

In considering the scaling limit, we want,
similarly to  \cite{HGSIV,HGSV}, to combine two seemingly inconsistent requirements: $\al\ne 0$ and $\rho(\theta)=\Rho(\theta)=1$. 
In fact this can be achieved for a discreet set of $\al$'s introducing
the fermionic screening operators \cite{HGSIV}, and then invoking the analytical continuation. 
As will be clear later our definition is consistent rather with the understanding on the model 
in terms of the action \eqref{action1}.

The scaling limit consists in taking in both Space and Matsubara directions staggering
inhomogeneietirs $\tau_j=(-1)^j\tau$, and considering the limit 
$$\tau\to \infty\,,\quad L\to \infty,\quad 2L  e^{-\tau}\to 2\pi MR\,\,\,\, \text{finite}\,,$$
where $R$ is the radius of the cylinder, $M$ is the mass of the soliton \eqref{mass}.

For $\rho(\theta)=\Rho(\theta)=1$ in the weak sense the operators $\tilde \bb^*(\z)$,  $\tilde \cb^*(\z)$  coincide
with the operators  $ \bb^*(\z)$,  $\cb^*(\z)$. Similarly to \cite{HGSIV, HGSV} the relations
\eqref{eq1} hint that the asymptotics for $\theta\to\pm\infty$ of the fermions (KM currents) 
are anti-periodic
(periodic) in $\theta$. Explicitly we assume
\begin{align}
&\bb^*(\theta)\ \ \simeq \hskip - .6cm{}_{{}_{{}_{{}_{\theta\to\pm\infty}}}}\sum\limits_{j=1}^\infty e^{\mp(2j-1)\theta}\bb^*_{2j-1}\,,\quad
\cb^*(\theta)\ \ \simeq \hskip - .6cm{}_{{}_{{}_{{}_{\theta\to\pm\infty}}}}\sum\limits_{j=1}^\infty e^{\mp(2j-1)\theta}\cb^*_{2j-1}\,,\nn\\
&\jb^\sigma(\theta)\ \ \simeq \hskip - .6cm{}_{{}_{{}_{{}_{\theta\to\pm\infty}}}}\sum\limits_{j=1}^\infty e^{\mp 2j\theta}\jb^\si_{2j}\,,\nn\quad \sigma=0,\pm\,.\nn
\end{align}
The Suzuki equations \eqref{eqbfinal} are obtained by this procedure from the corresponding lattice 
equations, which have the same structure as \eqref{eqbfinal}, but differ only by the driving term. In the case
of the lattice it is given by : 
\begin{align}
D(\theta)=2\sum_j\log\Bigl(\tanh\frac 1 2(\theta-\tau_j-i0)\Bigr)
-\frac{\pi i \nu\kappa}{1-2\nu}\,,\label{drive}
\end{align}
for which we have in the scaling limit
$$
D(\theta)\to -2\pi MR\cosh(\theta-\pi i \gamma)
-\frac{\pi i \sqrt{2}}{\beta}P\,.$$
The identification between $\kappa$ and $P$ is 
\begin{align}
 \sqrt{2} \beta P=\nu\kappa\,.\label{defP}
\end{align}

\subsection{Equations for $\Omega$}

Now we shall present a conjecture for the scaling limit of $\Omega(\theta,\theta')$ in the  case
$\al\ne 0$ and $\rho(\theta)=\Rho(\theta)=1$ and provide some justifications for it :
\begin{align}
\Omega(\theta,\theta')&=\int\limits_{-\infty}^{\infty}L(\theta-\eta+\pi i\gamma)\mathcal{G}(\eta-\pi i \gamma,\theta')dm_b(\eta-\pi i \gamma)\label{eqOm}\\&+\int\limits_{-\infty}^{\infty}L(\theta-\eta-\pi i\gamma)\overline{\mathcal{G}}(\eta+\pi i \gamma,\theta')d\overline{m}_b(\eta+\pi i \gamma)
\nn\,,\end{align}
where for the auxiliary functions we have the linear equations
\begin{align}
\mathcal{G}(\theta-\pi i \gamma,\theta')&=L(\theta-\theta'-\pi i \gamma)+\int\limits_{-\infty}^{\infty} L(\theta-\eta-\pi i \gamma)\Omega(\eta,\theta')dm_y(\eta)\label{eqG}\\&
+\int\limits_{-\infty}^{\infty}G_\al(\theta-\eta)\mathcal{G}(\eta-\pi i \gamma,\theta')dm_b(\eta-\pi i \gamma)\nn\\&-
\int\limits _{-\infty}^{\infty}G_\al(\theta-\eta+\pi i(1-2\gamma))
\overline{\mathcal{G}}(\eta+\pi i \gamma,\theta')d\overline{m}_b(\eta+\pi i \gamma)\,,\nn\end{align}
\begin{align}
\overline{\mathcal{G}}(\theta+\pi i \gamma,\theta')&=L(\theta-\theta'+\pi i \gamma)+\int\limits_{-\infty}^{\infty} 
L(\theta-\eta+\pi i \gamma)\Omega(\eta,\theta')dm_y(\eta)\label{eqbarG}\\&
+\int\limits_{-\infty}^{\infty}G_\al(\theta-\eta-\pi i(1-2\gamma))\mathcal{G}(\eta-\pi i \gamma,\theta')dm_b(\eta-\pi i \gamma)\nn\\&-
\int\limits _{-\infty}^{\infty}G_\al(\theta-\eta)\overline{\mathcal{G}}(\eta+\pi i \gamma,\theta')d\overline{m}_b(\eta+\pi i \gamma)\,,\nn
\end{align}
and we defined 
\begin{align}
&dm_y(\theta)=\frac{{y}(\theta)}{1+{y}(\theta)}\,,\quad dm_b(\theta)=\frac{{b}(\theta)}{1+{b}(\theta)}\,,\quad d\overline{m}_b(\theta)=\frac{\overline{b}(\theta)}{1+\overline{b}(\theta)}\,,
\nn\\
&L(\theta)=\frac {1}{2\pi \cosh\theta}\,,\quad
G_\al(\theta)=\frac 1 {4\pi}\int\limits _{-\infty}^{\infty}\frac 
{\sinh\(\frac{3\beta^2-1}{2(1-\beta^2)}{\pi k}+\frac{\pi i\al} 2\)}
{\sinh\(\frac{\beta^2}{1-\beta^2}{\pi k}+\frac{\pi i\al} 2\)\cosh\(\frac{1}{2}\pi k\)}
e^{ik \theta}dk\,.\nn
\end{align}
The shift $\gamma$ is an arbitrary real number from the interval  $(0,\pi/2)$.

The most important support for this definition is provided by the case $\al=0$ for which the
requirements $\rho(\theta)=\Rho(\theta)=1$ are automatic and do not require additional
work even on the lattice. In that case we have
\eqref{dy=O}
\begin{align}
&\frac{\delta}{\delta\tau}\log y(\theta)=\Omega(\theta,\tau+{\textstyle \frac{\pi i } 2})\,.\label{dy}
\end{align}
Using the Suzuki equations \eqref{eqbfinal} with the driving term replaced by $D(\theta)$ \eqref{drive}
one readily compute the variation with respect to any $\tau_j$ finding agreement with \eqref{eqOm}.
Strictly speaking even for $\al=0$ to  combine   the equations \eqref{dy} for all  $\tau_j$, we do not have
enough conditions to assert \eqref{eqOm} for all $\theta'$, but this is a very natural conjecture to make. 

The next question is how did we incorporate $\al$ into the equations \eqref{eqOm},  \eqref{eqG},
\eqref{eqbarG}. This was done due to the experience with equations of this kind
\cite{HGSIV,HGSV} and some meditation. Our choice is supported
by computation of the
residue at $\theta=\theta+\pi i$. After some rather tedious computation we obtain the 
following result
\begin{align}
\res\hskip - .8cm{}_{{}_{{}_{{}_{\theta=\theta'+\pi i}}}}\Omega(\theta, \theta')=\frac 1 {2\pi i}\frac{y(\theta')y(\theta'+\pi i)-1}{y(\theta')y(\theta'+\pi i)}\,,\nn
\end{align}
which coincides with the expected result from the definition \eqref{defOmega} and known singularities of $\omega(\theta,\theta')$
\cite{HGSIII}.
\subsection{Numerical results by interpolation}
\label{subsec:Num}
Our method of numerical investigation of the
equations \eqref{eqbfinal}  was explained 
in \cite{BaSm}. With these results at hand the numerical solution to the linear 
equations 
 \eqref{eqOm},  \eqref{eqG},
\eqref{eqbarG} 
is rather straightforward. The most interesting thing to study is the  limit $\theta_0\to\infty$ 
where we make contact with the UV CFT. We begin with the case $\theta\to\infty, \theta'\to \infty$
for which we assume
\begin{align}\Omega(\theta,\theta')\simeq\sum_{i,j=1}^\infty e^{-(2i-1)\theta}e^{-(2j-1)\theta'}D_{2i-1}(\al)D_{2j-1}(2-\al)\Omega_{2i-1,2j-1}(\theta_0)
 \,. \label{ExpOm}
\end{align}
The coefficients $D_{2i-1}(\al)$ are not hard to guess from \eqref{defC} and by analogy with \cite{HGSIV}:
\begin{align}
&D_m(\al)=i^m\sqrt{\frac\pi 2}\frac{\ \Gamma\(\frac m 2 \)\Gamma\(\frac 1 {1-\beta^2}m+\frac{\al}2\)}{(m-1)!\(\frac{m-1}2\)!\Gamma\(\frac {\beta^2}{(1-\beta^{2})}m+\frac{\al}2\)}\,.
\label{CoefD}
\end{align}
Additional support for this formula will be given below by considering the reflection relations. 
We have further
$$\lim_{\tth_0 \to \infty }e^{-2(i+j-1)\theta_0}\Omega_{2i-1,2j-1}(\theta_0)\to \Omega_{2i-1,2j-1}\,,$$
$\Omega_{2i-1,2j-1}$ is a polynomial of $P$ of degree $2i+2j-2$ with the leading coefficient equal to $1/(i+j-1)$.

We proceed with numerical checks of these assumptions. For $\theta_0=15$ we obtain already perfect
agreement with the scaling behaviour. The values of $P$ should not be to large, we take $P\le 0.2$. Considering an important amount
of numerical data with different $P,\al,\nu$ we come with the following conjectures for the
exact forms of the first several $\Omega_{m,n}$:
\begin{align}
\Omega_{1, 1} & = 
 P^2 - \frac{1}{16} - \frac{1}{8}\De_\al \,. \label{Om11} \\
2\cdot\Omega_{{1, 3}\atop{3,1}} &= 
 P^4 - 
 P^2\frac{ 5}{48} \(2\De_\al+3 \)  + 
 \frac{\hat{c}+8}{1536}(4\De_\al+3)
 + 
  \frac 1 {128} \De_\al^2\mp \frac{1}{96} d_\al \De_\al\,.\label{Om13}   
  \\
3\cdot\Omega_{3, 3}& =
 P^6 - 
  P^4 \frac 1 {64}(18 \De_\al+47) + 
  P^2\( \frac{27}{1280 } \De_\al^2+ 
     \frac{23 \hat{c}+378 }{3840} \De_\al + \frac{46 \hat{c}+881}{5120}\) \label{Om33}   
     \\&
      - \frac{1}{2048}\De_\al^3 - 
  \frac{40 \hat{c}+21 }{61440} \De_\al^2  - \frac{ 5 \hat{c}^2+52 \hat{c}+222  }{81920}(2\De_\al+3)
  \,.   
  \nn \\  
3\cdot\Omega_{{1,5}\atop{5,1}} & = 
 P^6 - 
  P^4 \(\frac{35}{48} + \frac{7}{24} \De_\al\) + 
  P^2 \( \frac{89}{3840 }\De_\al^2+ \frac{23 \hat{c}+514}{15360}(4\De_\al+5)
     \)  \label{Om15} 
     \\ &   - 
 \frac{1}{2048}\De_\al^3  - 
   \frac{ 10\hat{c}+479}{61440 }\De_\al^2
 - \frac{
  6 \hat{c}^2+83 \hat{c}+386}{245760}(6\Delta+5)  
    \nn \\&\mp
    d_\al\De_\al\( P^2\frac{23  }{960}   -  \frac{1}
     {512}\De_\al - \frac{83+12\hat{c}}{15360}\)\,. \nn
\end{align}

where
\begin{equation}
     d_\al{= \frac{1}{4}(\be^{-2}-\be^2)(\al-1)} \,.
     \label{dOdd}
    \end{equation}

Below we give some examples of comparison between numerical resuls and the
analytical conjectures above.
\vskip .2cm
\centerline{Coefficient $\Om_{1,1}$ and $ \beta^2 = \frac{1}{2}$ }
\vskip .2cm

\centerline{\scalebox{.9}{\begin{tabular}{| *{7}{c|} }
  \hline
      & \multicolumn{2}{c|}{$\al = 0.2$}
            & \multicolumn{2}{c|}{$\al = 0.4$}
                    & \multicolumn{2}{c|}{$\al = 0.6$}
                                           \\
  \hline
  $P$ & $\Om_{1,1}$ comp. & $\Om_{1,1}$ analyt. & $\Om_{1,1}$ comp. & $\Om_{1,1}$ analyt.& $\Om_{1,1}$ comp.& $\Om_{1,1}$ analyt.\\
  \hline 
0.02&-0.059287494 & -0.0592875&	-0.057099995&	-0.0571&	-0.055537495&	-0.0555375
\\
\hline
0.04&-0.058087495&	-0.0580875&	-0.055899995&	-0.0559&	-0.054337495&	-0.0543375
\\
\hline
0.06&-0.056087495&	-0.0560875&	-0.053899995	&-0.0539&	-0.052337495&	-0.0523375
\\
\hline
0.08&-0.053287495&	-0.0532875&	-0.051099995&	-0.0511&	-0.049537496&	-0.0495375
\\
\hline
0.1&-0.049687495&	-0.0496875&	-0.047499996&	-0.0475&	-0.045937496&	-0.0459375
\\
\hline
0.12&-0.045287496&	-0.0452875&	-0.043099996&	-0.0431&	-0.041537497&	-0.0415375
\\
\hline
0.14&-0.04008745&	-0.0400875&	-0.037899997&	-0.0379&	-0.03633745&	-0.0363375
\\
\hline
0.16&-0.034087497&	-0.0340875&	-0.031899998&	-0.0319&	-0.030337498&	-0.0303375
\\
\hline
0.18&-0.027287498&	-0.0272875&	-0.025099998&	-0.0251&	-0.023537499&	-0.0235375
\\
\hline
0.2&-0.019687499&	-0.0196875&	-0.017499999&	-0.0175&	-0.015937400&	-0.0159375
\\ 
\hline
\end{tabular}}}
\vskip .2cm
\centerline{Coefficient $\Om_{1,3}$ and $\beta^2 = \frac{3}{5}$ }
\vskip .2cm
\centerline{\scalebox{.9}{\begin{tabular}{| *{7}{c|} }
  \hline
      & \multicolumn{2}{c|}{$\al = 0.2$}
            & \multicolumn{2}{c|}{$\al = 0.4$}
                    & \multicolumn{2}{c|}{$\al = 0.6$}
                                           \\
  \hline
  $P$ & $\Om_{1,3}$ comp. & $\Om_{1,3}$ analyt. & $\Om_{1,3}$ comp. & $\Om_{1,3}$ analyt.& $\Om_{1,3}$ comp.& $\Om_{1,3}$ analyt.\\
  \hline
0.02&0.01612247&	0.01612249&	0.01591101&	0.01591102&	0.01577159&	0.01577160
\\ 
\hline
0.04&0.01575287&	0.01575289&	0.01554374&	0.01554376&	0.01540599&	0.01540600
\\ 
\hline
0.06&0.01514328&	0.01514329&	0.01493803&	0.01493804&	0.01480306&	0.01480307
\\ 
\hline
0.08&0.01430328&	0.01430329&	0.0141035&	0.01410349&	0.01397239&	0.01397240
\\ 
\hline
0.1&0.01324632&	0.01324633&	0.01305352&	0.01305353&	0.01292743&	0.01292744
\\ 
\hline
0.12&0.01198969&	0.01198969&	0.01180544&	0.01180545&	0.01168546&	0.01168547
\\ 
\hline
0.14&0.01055449&	0.01055449&	0.01038035&	0.01038036&	0.01026760&	0.01026760
\\ 
\hline
0.16&0.008965691&	0.008965693&	0.008803224&	0.008803226&	0.008698802&	0.008698804
\\ 
\hline
0.18&0.007252093&	0.007252093&	0.007102848&	0.007102848&	0.007007872&	0.007007871
\\ 
\hline
0.2&0.005446336&	0.005446333&	0.005311868&	0.005311867&	0.005227447&	0.005227444
\\ 
\hline
\end{tabular}}}
\vskip .2cm
\centerline{Coefficient $\Om_{3,3}$ and $\beta^2 = \frac{1}{2}$ }
\vskip .2cm
\hskip -.4cm\scalebox{.9}{\begin{tabular}{| *{7}{c|} }
  \hline
      & \multicolumn{2}{c|}{$\al = 0.2$}
            & \multicolumn{2}{c|}{$\al = 0.4$}
                    & \multicolumn{2}{c|}{$\al = 0.6$}
                                           \\
  \hline
  $P$ & $\Om_{3,3}$ comp. & $\Om_{3,3}$ analyt. & $\Om_{3,3}$ comp. & $\Om_{3,3}$ analyt.& $\Om_{3,3}$ comp.& $\Om_{3,3}$ analyt.\\
  \hline
0.02&-0.0079402716&	-0.0079402720&	-0.0078464501&	-0.0078464506&	-0.0077795388&	-0.0077795392\\ 
\hline
0.04&-0.0077381755&	-0.0077381759&	-0.0076463818&	-0.0076463822&	-0.0075809093&	-0.0075809097
\\ 
\hline
0.06&-0.0074059724&	-0.0074059727&	-0.0073175266&	-0.0073175270&	-0.0072544297&	-0.0072544302
\\ 
\hline
0.08&-0.0069505170&	-0.0069505174&	-0.0068666923&	-0.0068666926&	-0.0068068739&	-0.0068068743
\\ 
\hline
0.1&-0.0063812449&	-0.0063812453&	-0.0063032481&	-0.0063032484&	-0.006247564&	-0.0062475644
\\ 
\hline
0.12&-0.0057100111&	-0.0057100113&	-0.0056389640&	-0.0056389643&	-0.0055882093&	-0.0055882096
\\ 
\hline
0.14&-0.0049508823&	-0.0049508825&	-0.0048878029&	-0.0048878031&	-0.0048426982&	-0.0048426985
\\ 
\hline
0.16&-0.0041198841&	-0.0041198842&	-0.0040656674&	-0.0040656675&	-0.0040268458&	-0.0040268459
\\ 
\hline
0.18&-0.0032347010&	-0.0032347011&	-0.0031901003&	-0.0031901004&	-0.0031580934&	-0.0031580936\\ 
\hline
0.2&-0.0023143311&	-0.0023143311&	-0.0022799390&	-0.0022799391&	-0.0022551639&	-0.002255164
\\ 
\hline
\end{tabular}}
\vskip .2cm

The scaling limit of \eqref{defZ1} is supposed to give the ratio 
$$\frac{\langle\mathcal{O}_\al(0)\rangle_{P,R}}{\langle V_\al(0)\rangle_{P,R}}\,,$$
for some operator $\mathcal{O}_\al$. In the case under consideration 
this operator is supposed
to be a chiral descendant of $V_\al$ (recall that we do not distinguish between the CFT operators and
their perturbed counterparts). To be more precise $\Omega_{2i-1,2j-1}(\theta_0)$ should be related to
a descendant on the level $2i+2j-2$. The determinants made of  $\Omega_{2i-1,2j-1}(\theta_0)$
correspond to other descendants but we shall not discuss them here restricting ourselves to the simplest case. 

All together we must have 
\begin{align} \lim_{\theta_0\to \infty} e^{-2(i+i-1)\theta_0}\Omega_{2i-1,2j-1}(\theta_0)&=
 \frac{\langle \mathcal{P}_{2i-1,2j-1}(\{\mathbf{s}_k,\mathbf{l}_{2m}\})V_\al\rangle_{P}}{\langle V_\al(0)\rangle_{P}}\,,\label{desc}  
\end{align}
where $\mathcal{P}_{2i-1,2j-1}(\{\mathbf{s}_k,\mathbf{l}_{2m}\})V_\al$ is an element of the Verma module generated by $V_\al$ quotiented by the action of local
integrals of motion, this will be discussed in Section \ref{sec:Reflection}.

The expressions like the one in the right  hand side of \eqref{desc} can be computed for any $\mathcal{P}_{2i-1,2j-1}$,
we shall give some examples in the next section. However, trying to find $\mathcal{P}_{2i-1,2j-1}$ from \eqref{desc} 
we encounter more problems than in the usual Virasoro case \cite{HGSIV}. The point is that
the universal enveloping algebra of the super conformal algebra contains much more elements than that of the Virasoro algebra. The coefficients of the polynomials  $\mathcal{P}_{2i-1,2j-1}$ do not depend on $P$, and actually the
appearance of different degrees of $P$ is the source (the only one) of different equations. When the level grows the number of
coefficients of $\mathcal{P}_{2i-1,2j-1}$ grows much faster that the degree of the left hand side in $P$. For the Virasoro case we still could define the coefficients  up to the level $6$, and for levels $2$ and $4$ the systems of
equations were even overdetermined, the fact that they allowed solutions was considered as an important check 
of our procedure. In the super conformal case the only possibility to find the coefficients occurs on the level $2$: we
have two descendants created by $\mathbf{l}_{-2}$ and $\mathbf{s}_{-\frac 3 2}\mathbf{s}_{-\frac 1 2}$ and two coefficients of the polynomial in $P$ in the left hand side. Starting from the level $4$ we do not have enough equations.

One way out of this difficulty would be to allow descendants in the asymptotic states like it was done
in \cite{boos} 
for the level $8$ in the Virasoro case. This would be too hard, and not necessary: we have
another, similar to that of \cite{NeSm},  way of fixing the polynomials $\mathcal{P}_{2i-1,2j-1}$ based on the reflection relations \cite{FFLZZ,FLZZ1,FLZZ2}. We shall explain this in the next section. When the polynomials
$\mathcal{P}_{2i-1,2j-1}$ are defined from the reflection relations, the formulae 
\eqref{desc}, \eqref{Om11}, \eqref{Om13}, \eqref{Om15}, \eqref{Om33} 
can be used for checks. Since both our equation for $\Omega(\theta,\theta')$ and the reflection relations
have the status of conjectures the fact that the results of their application are in agreement
provides a very solid support for both.

\subsection{Primary fields} 
Let us now consider the asymptotics $\theta\to -\infty$, $\theta'\to \infty$. We have 
$$\Omega(\theta,\theta')\simeq\sum_{i,j=1}^\infty e^{(2i-1)\theta}e^{-(2j-1)\theta'}\Omega_{-(2i-1),2j-1}(\theta_0)\,.$$

We suspect that similarly to \cite{HGSV} the $\Omega_{-1,1}(\theta_0)$ is related to the ratio of the expectation
values of two shifted primary fields. The question is: which primary fields exactly? Now we have two of them:
$V_\al$, $W_\al$. Solving numerically our equations we find that for fixed $\beta, \al, P$
$$\log \Omega_{-1,1}(\theta_0)\simeq 2\theta_0\Bigl(\Delta_{\al+\frac{2\beta^2}{1-\beta^2}}+1/2-\Delta_\al
\Bigr)
\,.$$
Let us give an example. 
Consider the normalised expression:
$$R(\theta_0)= \exp\Bigl\{-2\theta_0\Bigl(\Delta_{\al+\frac{2\beta^2}{1-\beta^2}}+1/2-\Delta_\al\Bigr)\Bigr\}\Omega_{-1,1}(\theta_0)\,.$$
For $\al=1/2, \beta^2=1/2, P=0.1$ we have
\vskip .3cm
\centerline{{\begin{tabular}{|c|c|c|c|c|c|c|c|c|}
\hline
$\theta_0$
&12& 13 &  14 &  15 &  16\\
\hline
$R(\theta_0)$&0.16825979& 0.16825580&  0.16825433& 
0.16825379& 0.16825359\\
\hline
\end{tabular}}}
\vskip .3cm
So, we see that the scaling is achieved with great precision. 

This suggests that $ \Omega_{-1,1}(\theta_0)$ is proportional to the ratio of the expectation values
of $W_{\al+\frac{2\beta^2}{1-\beta^2}}$ and $V_\al$. 
Let us check the limiting value against the CFT. First, we have to normalise the primary fields
$$\hat{V}_{\al}=\frac 1 {F(\al)}{V}_{\al}\,,\quad \hat{W}_{\al}=\frac 1 {F(\al)}{W}_{\al}\,,$$
where $F(\al)$ is the one-point function of the operator ${V}_{\al} $ on the plane (for $R=\infty$) \cite{InfVol}.
For the operator ${W}_{\al}$ the one-point function on the plane vanishes since this operator
is a super Poincare descendant of ${V}_{\al} $ and the vacuum is super Poincare invariant. 
Nevertheless we normalise ${W}_{\al}$ by the same function $F(\al)$. The reason for that is
in the reflection relations as explained in the next section. 
Denote by $c(\al,P)$ ($\tilde c(\al,P)$) the CFT one-point functions of the normalised
operator $\hat{V}_{\al}$ ( $\hat{W}_{\al}$) on the cylinder with our usual asymptotic conditions. 
In the next section we compute
\begin{align}
\frac{\tilde{c}(\al+\frac{2\beta^2}{1-\beta^2},P)}{c(\al,P)}&=\frac{\pi ^2} {1-\beta^2}\beta^{ \frac 1 2(\al\beta^2-2\beta^2-\al)} \frac{\gamma(\frac 1 2(1-\beta^2)(2-\al ) ) }{\gamma(\frac 1 4(1-\beta^2) (2-\al ) )^2} \label{cbyc}\\&\times\gamma(
    \half (1+\beta^2) + (1-\beta^2) \al - \beta P) \gamma(
     \half (1+\beta^2) +  
    (1-\beta^2) \al +\beta P ) \,.\nn
\end{align}
Consider the ratio
\begin{align}
R_1(\theta_0)=R(\theta_0)\frac{c(\al,P)}{\tilde{c}(\al+\frac{2\beta^2}{1-\beta^2},P)}\,.
\end{align}
For $\theta_0=15$ and a random choice of $\nu,\al,P$ we have
\vskip .3cm
\centerline{\begingroup
\setlength{\tabcolsep}{10pt} 
\renewcommand{\arraystretch}{1.5} 
{\begin{tabular}{|c|c|c|c|}
\hline
data
& $\beta^2=\frac1 2,\al=\frac 2 5,P=0.2$& $\beta^2=\frac 35,\al=\frac 2 3,P=0.1$& $\beta^2=\frac1 3,\al=\frac 1 2,P=0.15$\\ 
\hline
$R_1(15)$&1.00000211&1.00009870&0.99999998\\
\hline
\end{tabular}
}\endgroup}

\vskip .3cm
The agreement is very good.

\section{Reflection relations and three-point functions in super CFT}
\label{sec:Reflection}
Long ago Al. Zamolodchikov did a remarkable observation that the one-point functions
for sine-Gordon and sinh-Gordon model are related by analytical continuation. This is
very different from other properties of these models, for example the particle content
is quite different. Nevertheless the Al. Zamolodchikov's observation proved to be 
correct in many cases. Here we shall apply it to the ssG model relating it to the super sinh-Gordon
with the action 
\begin{align}
\mathcal{A}_\mathrm{L}'=
\int\Bigl[\Bigl( \frac 1 {4\pi}\partial_z\varphi\partial_{\bar{z}} \varphi+
\frac 1 {2\pi}\left(\psi\partial_{\bar z}\psi+\bar{\psi}\partial_z\bar{\psi}\right)&-\mu \bar{\psi}\psi e^{\frac{b}{\sqrt{2}}\varphi}\Bigr)-
\mu \bar{\psi}\psi e^{-\frac{b}{\sqrt{2}}\varphi}\Bigr]d^2z\,.
\label{action2}
\end{align}
We shall use the habitual notation
$$Q=b+b^{-1}\,.$$
The analytical continuation to the ssG case corresponds to
\begin{align}
\beta=ib\,,\qquad \al=\frac{2a}Q\,.\label{constants}
\end{align} 
{Slightly abusing the notation we will write the primary fields \eqref{VWDef} as $V_a$ and $W_a$.}
The idea behind the reflection relations is that the physical quantities must be
invariant under the two reflections:
\begin{align}
\si_1 \, : \, a \to -a \,, \quad\quad \si_2 \, : \, a \to Q-a \,,\label{reflections}
\end{align}
The first of them reflects simply the $C$-reflection of the action \eqref{action2} while
the second one is inherited from the symmetry of the super Liouville model. 
{The reflection relations can be applied to the calculation of one point functions. For the primary fields it is rather direct, since their one point functions
are invariant under $\si_1$ and their transformation rule under $\si_2$ is inherited from a remarkable property of the (super) Liouville three point function.
This will be explained in more details in Section \ref{subsec:Primary}. 
The situation is more complicated for descendants fields : a Virasoro descendants has a manifest $\si_2$ symmetry, but its behaviour for $\si_1$ is unclear.
This explains the necessity to construct a passage matrix $U(a)$ that relates the Virasoro and Heisenberg descendants in order to use the
action of the two reflections simultaneously. Recall that $\mathcal{V}^\mathrm{quo}_a$ is the quotient of the Verma module by the action of the local integrals of motion. 
Consider $V(a)\in \(\mathcal{V}^\mathrm{quo}_a\)^*$. The reflection relations \cite{FFLZZ} can be presented as the
following Riemann-Hilbert problem (see \cite{NeSm} for more details):
\begin{equation}
V(a+Q) = S(a) V(a)  \,, \quad S(a)  = U(-a)U(a)^{-1}  \,.
\end{equation}}
Let us apply this idea. 
\subsection{Primary fields}
\label{subsec:Primary}
We begin with the primary fields for which the three-point function
on a sphere on the one hand and the one-point function on the cylinder (with our usual asymptotical conditions) on the other hand 
coincide. The main reference for this subsection is \cite{SuperL}.
Consider the three-point function
of the fields $V_{a_j}$, $j=1,2,3$:
\begin{align}
C(a_1,a_2,a_3)&=\(\half  \pi\mu\gamma(\half bQ)b^{-1-b^2}\)^{\frac{Q-a} b}\YN(2a_1)\label{3p}
\\&\times\frac{\YN'(0)\YN(2a_2)\YN(2a_3)}{\YN(a-Q)\YN(a_{1+2-3}))\YN(a_{2+3-1}))\YN(a_{3+1-2})}\,, \nn
\end{align}
where $a=a_1+a_2+a_3$, $a_{1+2-3}=a_1+a_2-a_3$ {\it etc},
\begin{align}
\YN(x)=\Upsilon\(\frac{x} 2\)\Upsilon\(\frac{x+Q} 2\)\,,\qquad 
\YR(x)=\Upsilon\(\frac{x+b} 2\)\Upsilon\(\frac{x+b^{-1}} 2\)\,,\nn
\end{align}
and the well-known function $\Upsilon(x)$ satisfies the identities
\begin{align}
&\frac{\Upsilon(x-b)}{\Upsilon(x)}=b^{2bx-2b^2-1}\gamma(1+b^2-bx)\,,\quad
\frac{\Upsilon(x-b^{-1})}{\Upsilon(x)}=\gamma(1+b^{-2}-xb^{-1})b^{1-2x/b+2b^{-2}}\,.\nn\\
&\Upsilon(Q-x)=\Upsilon(x)\,.\nn
\end{align}
{We recall that $\ga(z) = \frac{\Ga(z)}{\Ga(1-z)}$. One can use the following integral representation for $\log\Up$ 
in the range $0<\mathrm{Re}(x)<Q$ :
\begin{equation}
\log\Upsilon(x)= \int_0^\infty \frac{dt}{t} \bigg\{
\left( \frac{Q}{2}-x \right)^2 e^{-t} - \frac{\ssh^2\left((\frac{Q}{2}-x)\frac{t}{2}\right)}{\ssh(\frac{tb}{2})
\ssh(\frac{t}{2b})}
\bigg\}\,.    \label{UpInt}  
\end{equation}
}

The function $\YR(x)$ was introduced in order to be able to write down the three-point
function for the fields $W_{a_1},V_{a_2},V_{a_3}$:
\begin{align}
\widetilde{C}(a_1,a_2,a_3)&=\(\half  \pi\mu\gamma(\half bQ)b^{-1-b^2}\)^{\frac{Q-a} b}\YN(2a_1)\label{3ptilde}
\\&\times\frac{2i\YN'(0)\YN(2a_2)\YN(2a_3)}{\YR(a-Q)\YR(a_{1+2-3}))\YR(a_{2+3-1}))\YR(a_{3+1-2})}\,.  \nn
\end{align}

In the formulae \eqref{3p}, \eqref{3ptilde} we separated the multiplier in the first line
from the rest because this is the only one which is not invariant under $a_1\to Q-a_1$. 
This gives the possibility   to compute the reflection coefficient relating 
\begin{align}
&V_a=R(a)V_{Q-a}\,,\quad W_a=R(a)W_{Q-a}\,,\nn\\
&R(a)=\bigl(\pi \mu \gamma(b^2)\bigr)^{\frac{Q-2a} b}b^{-2}\gamma(2ab-b^2)\gamma(2ab^{-1}-b^{-2}-1)\,.\nn
\end{align}
The one-point function of $V_a$ in infinite volume  for super sinh-Gordon 
is invariant under both reflections \eqref{reflections}, hence it satisfies
$$F(a)=F(a-Q)R(a)\,.$$
The operators
$$\hat{V}_a=\frac 1 {F(a)}V_a\,,\qquad \hat{W}_a=\frac 1 {F(a)}W_a\,,$$
are invariant under both reflections. 
For our goals we do not need $F(a)$ but rather the ratio $f(a)=\frac{F(a-b)}{F(a)}$ for which
$$f(a-Q)=f(a)\frac{R(a)}{R(a-b)}\,.$$
We compute and rewrite the result in a useful for us way
\begin{align}
&\frac{R(a)}{R(a-b)}
=\(\half \pi\mu\gamma(\half bQ)\)^{-2}\frac{\gamma\(\half+\half b(2a-b)\)
}{
\gamma\(\frac 1 2 +\half b(2(a-Q)-b)\)
}\,.\nn
\end{align}
This equality implies
\begin{align}
f(a)
=C(b)\(\half \pi\mu\gamma(\half bQ)\)^{\frac2 {bQ}(\Delta_{a-b}+\frac 1 2-\Delta_a)}\gamma\(\half b( Q-2a)\)\,,\label{f}
\end{align}
where $C(b)$ is a constant depending on $b$ only.
To finish the consideration of primary fields let us we give the expression for the
ratio
\begin{align}
\frac{\widetilde{C}(a-b,Q/2+k,Q/2-k)}{C(a,Q/2+k,Q/2-k)}&=
\(\half  \pi\mu\gamma(\half bQ)b^{-1}\)\frac{\gamma^2(\frac 1 2 (1+ab-b^2))\gamma(\frac 1 2b(Q-2a))}
{\gamma (ab-b^2)}\label{CbyC}\\ &\times\gamma(\half(1-b^2+ab)+bk)\gamma(\half(1-b^2+ab)-bk)\,.\nn
\end{align}
Divide  \eqref{CbyC} by $f(a)$ \eqref{f} (there is an important cancelation) and change the variables
by \eqref{constants}
and
$$bk=\beta P\,,$$
after some simplification this gives \eqref{cbyc}.

\subsection{Descendants}
\label{subsec:Descendants}
Here we find it more convenient to begin with the one-point functions for CFT on the cylinder. 
This is not absolutely trivial for the descendants. 
In the case of Super CFT, one should consider both descendants created by the stress energy tensor $T$ and by the super current $S$. 
Denote
\begin{equation}
\braket{\bl_{-n_1}...\bl_{-n_p}\bs_{-r_1}...\bs_{-r_q} V_a (y) } = 
\frac{\braket{\De_-|\bl_{-n_1}...\bl_{-n_p}\bs_{-r_1}...\bs_{-r_q} V_a(y)|\De_+ }}{\braket{\De_-|V_a(y)|\De_+}}  \,,
\label{GenVev}
\end{equation}
where $\De_\pm$ are primary states, located at the extremities $\pm \infty$ of the cylinder, and $\bl_m,\bs_r$ are defined in \eqref{ActionTS}.
The main idea is to follow the route of \cite{HGSIV}, where 
Ward-Takahashi identities have been used to obtain the values of the same kind of correlation functions but containing purely Virasoro generators. 
Using Ward-Takahashi identities to express the correlation functions $\braket{T(z_1)...T(z_p)S(w_1)...S(w_q)V_a(y)}$ we can then obtain \eqref{GenVev} by 
a successive application of \eqref{ActionTS} :
\begin{align}
\braket{\bl_{-n_1}...\bl_{-n_p}\bs_{-r_1}...\bs_{-r_q} V_a } & = 
\oint_{c_{z_1}}\boldsymbol{dz_1}...\oint_{c_{z_p}}\boldsymbol{dz_p}\oint_{c_{w_1}}\boldsymbol{dw_1}...\oint_{c_{w_q}}\boldsymbol{dw_q} \label{IntVev} \\
& \times \braket{T(x_1)...T(x_p)S(w_1)...S(w_q)V_a(y)}  \,, \nn
\end{align}
with the notation :
\begin{equation}
\oint_{c_{z_k}}\boldsymbol{dz_k} = \oint_{c_{z_k}}\frac{dz_k}{2\pi i (z_k-y)^{n_k-1}} \,, \quad\quad 
\oint_{c_{w_j}}\boldsymbol{dw_j} = \oint_{c_{w_j}}\frac{dw_j}{2\pi i (w_j-y)^{r_j-\frac{1}{2}}} \,,
\end{equation}
and the contours being small concentric circles around the point $y$ : $c_{z_1} \subset ... \subset c_{w_q} $. 
Using the OPEs \eqref{OPE}, \eqref{OPEprimary},  we can deduce the following simple correlation functions between the fields (we present here only the specific
identities that we shall need later) : 
\begin{align}
\braket{ S(x)V_a(y) }  = & 0  \nn  \,, \\
\braket{ S(x)W_a(y) }  = &\left( 2 \De_a \xi'(x-y)   - \xi(x-y)(\De_+-\De_-) \right)  \langle V_a (y) \rangle \nn \,, \\
\braket{ S(x_2) S(x_1)V_a(y) } =&  -\xi(x_1-y) \braket{S(x_2) W_a(y) } -2 \xi(x_1-x_2) \braket{T(x_2)V_a (y)}
+
\nn \\
& -\frac{c}{3}\xi''(x_1-x_2) \braket{V_a(y)}  \nn \,. 
\end{align}
And the more complicated : 

\begin{align}
& \braket{ T(x_3) S(x_2) S(x_1)V_a(y) }  = \nn \\
& \left( -\frac{3}{2}\chi'(x_3-x_2)  + (\chi(x_3-x_2)-\chi(x_3-y)) \frac{\pa}{\pa x_2}
 \right)  \braket{ S(x_2) S(x_1)V_a(y) } \nn \\
 & +\left( -\frac{3}{2}\chi'(x_3-x_1)  + (\chi(x_3-x_1)-\chi(x_3-y)) \frac{\pa}{\pa x_1}
 \right)  \braket{ S(x_2) S(x_1)V_a(y) } \nn \\
 &+
 \left( -\De_a \chi'(x_3-y)  + \chi(x_3-y) (\De_+-\De_-) + \left(\frac{\De_+ +\De_-}{2}-\frac{c}{24}\right)
 \right)  \braket{ S(x_2) S(x_1)V_a(y) } \nn  \,,
\\
&  \braket{ T(x_4) T(x_3) S(x_2) S(x_1)V_a(y) }  = \nn \\ & 
 -\frac{c}{12}\chi'''(x_4-x_3)  \braket{  S(x_2) S(x_1)V_a(y) } \nn \\ & +
  \left( -2\chi'(x_4-x_3)  + (\chi(x_4-x_3)-\chi(x_4-y)) \frac{\pa}{\pa x_3}
 \right)  \braket{ T(x_3) S(x_2) S(x_1)V_a(y) } \nn \\ &
 + \left( -\frac{3}{2}\chi'(x_4-x_2)  + (\chi(x_4-x_2)-\chi(x_4-y)) \frac{\pa}{\pa x_2}
 \right)  \braket{ T(x_3) S(x_2) S(x_1)V_a(y) } \nn \\
 & +\left( -\frac{3}{2}\chi'(x_4-x_1)  + (\chi(x_4-x_1)-\chi(x_4-y)) \frac{\pa}{\pa x_1}
 \right)  \braket{ T(x_3) S(x_2) S(x_1)V_a(y) } \nn \\
 &+
 \left( -\De_a \chi'(x_4-y)  + \chi(x_4-y) (\De_+-\De_-) + \left(\frac{\De_+ +\De_-}{2}-\frac{c}{24}\right)
 \right)  \braket{T(x_3) S(x_2) S(x_1)V_a(y) } \nn
\end{align}
As has been explained we apply this formulae in the case of equal boundary conditions $\De_+ = \De_- = \delta_P$.
The calculation of one point functions of descendants on the cylinder  are then given by the application of \eqref{IntVev}.
As examples we present the results at level 2 
\begin{equation}
\braket{\bl_{-2}V_a} = \de_P - \frac{c}{24} - \frac{\De_a}{12} \, ,
\quad\quad 
\braket{\bs_{-\frac{3}{2}}\bs_{-\frac{1}{2}}V_a} = \frac{\De_a}{12} \,, \label{1PFLvl2}
\end{equation}
and at level 4  : 
\begin{align}
& \braket{\bl^2_{-2}V_a} = 
\( \de_P - \frac{c}{24} \)^2 -\frac{1}{6} \( \de_P - \frac{c}{24} \)  -\frac{\De_a}{6}\( \de_P - \frac{c}{24} \) +\frac{\Delta^2_\alpha}{144}+\frac{7}{360}\De_a
 \,,  \label{1PFLvl4}
\\
&\braket{\bs_{-\frac{7}{2}}\bs_{-\frac{1}{2}}V_a} = - \De_a  \frac{7}{960} \,,
\quad \quad
\braket{\bs_{-\frac{5}{2}}\bs_{-\frac{3}{2}}V_a}
=
-\frac{1}{12}\( \de_P - \frac{c}{24} \) + \De_a \frac{17}{960} + \frac{7c}{2880} \,, \nn
\\
&\braket{\bl_{-2}\bs_{-\frac{3}{2}}\bs_{-\frac{1}{2}}} = 
\frac{\De_a}{12} \left(\de_P - \frac{c}{24}- \frac{\de_a}{12} \right) - \frac{1}{144}\De_a \,,
\quad\quad
 \braket{\bl_{-4}V_a} = \frac{\De_a}{240} \,.  \nn
\end{align}
We also will need the one point functions at level 6. Since the results are quite long, we prefer to display them in due time.

\subsection{Super Virasoro and super Heisenberg algebras}
We would like to have an independent check of the results \eqref{Om11} - \eqref{Om15}. 
In order to do so,  we should intepret  the expressions obtained for 
$\bbe_{2m-1}^{*}\bga_{2m-1}^{*}V_a $  as  decompositions of the fermionic
operators on the Super Virasoro basis, and check that this decomposition is compatible with the reflection relations.
As has been explained above and is clear from the interpretation of the reflections, it is first important to make the connection between the Super-Virasoro and
the Super-Heisenberg algebras, that is to construct the passage matrix $U(a)$. This is our goal in this subsection. 
%
\\
\\
The expression of the stress energy tensor and the super current in terms of  the
fields in the action \eqref{action2}  are given by :
\begin{align}
T(z) & = - \frac{1}{4} (\pa_z \varphi)^2  + \frac{Q}{2\sqrt{2}} \pa^2\varphi -\frac{1}{2} \psi\partial\psi \,,\nn \\
S(z) & =  i \( \frac{1}{\sqrt{2}} \psi \partial \vp -   Q \partial \psi\)  \,.\nn
\end{align}
In order to exhibit  the Heisenberg basis, we split  the field $\varphi(z,\bar{z})=\phi(z)+\phi(\bar{z})$ in chiral 
parts and expand in modes : 
\begin{equation}
\phi(z) = \phi_0 -2 i \pi_0 + i \sum_{k\in\mathbb{Z}^*} \frac{a_k}{k} z^{-k} \,,\nn
\end{equation}
where the  Heisenberg algebra is :
\begin{equation}
[a_k , a_l] = 2 k \delta_{k,-l} \,, \quad\quad [\phi_0,\pi_0] = i \,. 
\label{HAlg}
\end{equation}
The same analysis holds for the fermionic field
\begin{equation}
\psi (z) =  \sum_{r\in\mathbb{Z}} b_{r+\frac{1}{2}} z^{-r-1} \,,
\nn
\end{equation}
with the fermionic algebra defined by : 
\begin{equation}
\{ b_r , b_s \}  = \delta_{r,-s} \,.
\label{FAlg}
\end{equation}
We will call the combination of \eqref{HAlg} and \eqref{FAlg} the super Heisenberg algebra (together with the commutation relation
$[a_k,b_r]=0$).
The primary field $e^{\frac{a}{\sqrt{2}}\phi(0)}$ is identified with the highest weight vector of the super Heisenberg algebra : 
\begin{equation}
e^{\frac{a}{\sqrt{2}}\phi(0)} \Longleftrightarrow e^{\frac{a}{\sqrt{2}}\phi_0} \ket{0} \,, \quad a_k \ket{0} = b_r \ket{0} = 0 \,, \quad k,r >0 
\nn
\end{equation}
In the general case we should then take : 
\begin{equation}
V_a =  e^{\frac{a}{\sqrt{2}}(\phi_0+\overline{\phi_0})} \ket{0}\otimes \overline{\ket{0}}  \,.
\end{equation}
The calculation for the two chiralities being independent, we will work only with the holomorphic one.
We can now introduce the generators of the super Virasoro algebra : 
\begin{align}
 \bl_m =& 
\frac{1}{4}\sum_{k\neq 0,m}: a_k a_{m-k}: +  (\pi_0^2+i \pi_0 \frac{Q}{\sqrt{2}}) \delta_{m,0}+\nn
\\  & 
(\pi_0+i \frac{Q}{2\sqrt{2}} (m+1)) a_m(1-\delta_{m,0})
+
\frac{1}{2} \sum_{k\in \widetilde{\mathbb{Z}}} : b_{m-k} b_{k}: (k+\frac{1}{2}) \,, \nn
\end{align}
and the modes of the super current : 
\begin{equation}
\bs_{r} =
 \frac{1}{\sqrt{2}}\sum_{k\in \widetilde{\mathbb{Z}}'}
 b_ka_{r-k}  + \left( \sqrt{2}  \pi_0 +    i  Q (r+\frac{1}{2})\right)b_r  \,.\nn
\end{equation}
Here the symbol $:...:$ means 
normal order. 
These generators satisfy the super Virasoro algebra 
\begin{align}
    [\bl_m,\bl_n] & =(m-n)\bl_{m+n}+\frac{c}{12}m(m^2-1)\delta_{m,-n},  \nn\\
    \{\bs_r,\bs_s \} & = 2 \bl_{r+s}+\frac{c}{3}(r^2-\frac{1}{4})\delta_{r,-s}, \nn
\end{align}
with $ c =\frac{3}{2}(1+2Q^2) $, and since  $S$ is a primary field of conformal dimension $\Delta = \frac{3}{2}$ we also have the relation : 
\begin{equation}
[\bl_m,\bs_r]=    \left( \frac{m}{2}-r\right) \bs_{m+r}  \,.\nn
\end{equation}
Finally, the natural identity holds  : 
\begin{equation}
\bl_0 V_a = \De_a V_a \,, \quad\quad \De_a = \frac{1}{2}a(Q-a)  \,.\nn
\end{equation}
We are now ready to compute the passage matrix between the Super-Virasoro and the super Heisenber basis. Recall that we work modulo the action of 
local integrals of motion. For our calculations (up to level 6), the integrals of motion that will be involved are just the first two given by the 
densities \eqref{densities}. Explicitly : 
\begin{align}
\bi_1 & = \bl_{-1} \,, \label{i1} \\
\bi_3 & = 2 \sum_{k=-1}^\infty \bl_{-3-k}\bl_k + \frac{1}{2}  \sum_{k=-\frac{1}{2}}^\infty \bs_{-3-k}\bs_k \left( k + \frac{3}{2} \right)\,. \label{i3}
\end{align}

\subsubsection{Level 2.}
At level 2 there is only one integral of motion to take into account : 
\begin{equation}
\bi_1^2 V_a =  \bl_{-1}^2 V_a = 0  \,.\nn
\end{equation}
We define $U^{(2)}$ to be the passage matrix between the base $\{ \bl_{-2},\bs_{-\frac{3}{2}}\bs_{-\frac{1}{2}} \}$ and
$\{ a_{-1}^2,b_{-\frac{3}{2}}b_{-\frac{1}{2}} \}$ which is found to be : 
\begin{equation}
U^{(2)}=\left(
\begin{array}{cc}
 \frac{1}{4} \left(2 a^2+Q a+1\right) & \frac{1}{2} \\
 \frac{a^2}{2} & -a (a+Q) \\
\end{array}
\right) \,. \nn \label{ULvl2}
\end{equation}
Its determinant  factorises and gives as expected the null-vector conditions  :
\begin{equation}
\det (U^{(2)}) = -\frac{1}{4} a \left(2 a+{b}+{b^{-1}}\right)(a+b)(a+b^{-1})  \,.\nn
\end{equation}

\subsubsection{Level 4.}
\label{subsubsec:Lvl4}
At this level there are 10  operators in total, but working modulo integrals of motion (in this case also only $\bi_1$)
we need to keep only 5 of them, that we choose to  be
\begin{equation}
\bl_{-2}^2\,, \quad\quad \bl_{-4}\,, \quad\quad
\bs_{-\frac{7}{2}}\bs_{-\frac{1}{2}}\,,  \quad\quad \bs_{-\frac{5}{2}}\bs_{-\frac{3}{2}}\,, \quad\quad
\bl_{-2}\bs_{-\frac{3}{2}}\bs_{-\frac{1}{2}} \,. \nn
\end{equation}
On the other hand, we select the following operators to describe the states at level 4 from the super Heisenberg algebra point of view : 
\begin{equation}
a_{-2}^2\,,   \quad a_{-3}a_{-1} \,, \quad\quad b_{-\frac{7}{2}}b_{-\frac{1}{2}} \,, \quad b_{-\frac{5}{2}}b_{-\frac{3}{2}} \,,
\quad a_{-1}^2 b_{-\frac{3}{2}}b_{-\frac{1}{2}} \,.  \nn
\end{equation} 
We find for the matrix $U^{(4)}$ :
\begin{equation}
U^{(4)} = \left(
\begin{smallmatrix}
 U^{(4)}_{11} & U^{(4)}_{12} & \frac{29 a+12 Q}{4 a} &
   \frac{5 a+4 Q}{4 a} & \frac{1}{4} \left(2 a^2+Q a+1\right) \\
 \frac{1}{4} & \frac{1}{12} \left(2 a^2+3 Q a+6\right) & \frac{3}{2} & \frac{1}{2} & 0 \\
 0 & \frac{a^2}{6} & -a^2-3 Q a+3 & 1 & \frac{a^2}{2} \\
 0 & \frac{1}{6} \left(a^2+Q a+3\right) & -\frac{4 (2 a+3 Q)}{a} & U^{(4)}_{44} & \frac{1}{2}
   \left(-a^2-Q a+1\right) \\
 \frac{1}{4} \left(-2 a^2-Q a-1\right) & \frac{1}{6} \left(2 a^2-3\right) & U^{(4)}_{53} &
   U^{(4)}_{54} & U^{(4)}_{55} \\
\end{smallmatrix}  
\right) \label{U4}  \,,
\end{equation} 
where the lengthiest coefficients are : 
\begin{align}
U^{(4)}_{11} & = -\frac{ 4 a^4+4 Q a^3+Q^2 a^2+4 a^2+2 Q a+3}{8 a^2} \,, \quad U^{(4)}_{12}  = \frac{4 a^4+2 Q a^3-6 a^2-6 Q a-9}{12 a^2} \,,  \nn \\ 
U^{(4)}_{44} & =  -\frac{a^3+3 Q a^2+2 Q^2 a+3 a+4 Q}{a} \,, \quad  U^{(4)}_{53}  = \frac{1}{2} \left(-17 a^2-23 Q a-6 Q^2+6\right)  \,, \nn \\
U^{(4)}_{54} & = \frac{1}{2} \left(-a^2-3 Q a-2 Q^2+2\right)   \,, \quad  U^{(4)}_{55}  = -\frac{1}{4} a \left(2 a^3+3 Q a^2+Q^2 a+Q\right)  \,.  \nn
\end{align} 

Its determinant can be factorised  : 
\begin{equation}
\det(U^{(4)}) = \frac{1}{384} \frac{D^{(4)}_V(\Delta,c)}{D^{(4)}_H(a^2,Q^2)}N^{(4)}(a,b)  \,. \nn
\end{equation}
The contribution from the null vectors is :
\begin{align}
N^{(4)}(a,b) & = a^4 (a+b)^2(a+b^{-1})^2(a+2b)(a+2b^{-1})(a+3b)(a+3b^{-1}) \\ & \times (2a+b+b^{-1})
(2a+b+3b^{-1})(2a+3b+b^{-1})  \nn \,,
\end{align}
and we have :
\begin{equation}
D^{(4)}_V(\Delta,c) = 1 \,, \quad\quad \quad  D^{(4)}_H(a^2,Q^2) =a^2 \,. \label{detU4}
\end{equation}

\subsubsection{Level 6.}
We proceed through the same analysis. At level 6 we will need to factor out the action of  both $\bi_1$ and $\bi_3$.
There are 28 Virasoro operators at level 6, but the factorisation of the action of the integrals of motion leaves only 10, that we choose to be : 
\begin{align}
&\bl_{-2}^3\,, \quad \bl_{-6}\,, \quad \bl_{-3}^2\,, \quad  \bs_{-\frac{7}{2}} \bs_{-\frac{5}{2}}\,,  
\quad  \bs_{-\frac{9}{2}} \bs_{-\frac{3}{2}}\,, \quad  \bs_{-\frac{11}{2}} \bs_{-\frac{1}{2}} \,,\nn\\
& \bl_{-2}^2\bs_{-\frac{3}{2}}\bs_{-\frac{1}{2}}\,, \quad \bl_{-2}\bs_{-\frac{7}{2}}\bs_{-\frac{1}{2}}\,, 
\quad \bl_{-2}\bs_{-\frac{5}{2}}\bs_{-\frac{3}{2}}\,, \quad \bl_{-3}\bs_{-\frac{5}{2}}\bs_{-\frac{1}{2}} \nn \,.
\end{align}
These are expressed on the super Heisenberg basis :
\begin{align}
& a_{-1}^6\,, \quad a_{-1}^4a_{-2}\,, \quad a_{-3}^2\,, \quad b_{-\frac{7}{2}}b_{-\frac{5}{2}}\,, \quad b_{-\frac{9}{2}}b_{-\frac{3}{2}}\,,
\quad b_{-\frac{11}{2}}b_{-\frac{1}{2}} \,, \nn \\
& a_{-1}a_{-2} b_{-\frac{5}{2}}b_{-\frac{1}{2}}\,, \quad a_{-1}^2 b_{-\frac{7}{2}}b_{-\frac{1}{2}}\,, \quad a_{-1}a_{-3} b_{-\frac{3}{2}}b_{-\frac{1}{2}} \,,
\quad a_{-1}^2 b_{-\frac{5}{2}}b_{-\frac{1}{2}} \,.\nn 
\end{align}
The passage matrix $U^{(6)}$ is to big to be presented here, but we can give its determinant : 
\begin{align}
 \det(U^{(6)}) = -\frac{1}{212336640 } N^{(6)}(a,b)\frac{D_V^{(6)}(\De,c)}{D_H^{(6)}(a^2,Q^2)}  \,.
\end{align}
with :
\begin{align}
 N^{(6)}(a,b) & = a^2 (a+b)^5 (a +b^{-1})^5   (a+2 b)^2 (a+3 b)^2 (a +2b^{-1})^2 (a +3b^{-1})^2    \\
 & \times (a+4 b) (a+5 b) (a +4b^{-1}) (a +5b^{-1}) \nn\\
 & \times \left(a +b+b^{-1}\right) \left(2 a +b+b^{-1}\right)^5 \left(2 a +b+3b^{-1}\right)^2 \left(2 a +3 b+b^{-1}\right)^2 \nn\\
 & \times \left(2 a +5 b+b^{-1}\right) \left(2 a +b+5b^{-1}\right) \,, \nn 
\end{align}
the null vector contribution, and 
\begin{equation}
D_H^{(6)}(a^2,Q^2) =a^2 (-15 + 3 a^2 - 10 Q^2) \,,
\quad\quad
D_V^{(6)}(\De,c) = 1  \,.  \label{detU6}
\end{equation}

\subsection{Reflections relations}
We claim that similarly to \cite{NeSm}, the action of both reflections $\si_1$ and $\si_2$ implies that the fermions transform as : 
\begin{equation}
\bbe^{*}_{2j-1} \to \bga^*_{2j-1} \,,\quad \bga^*_{2j-1} \to \bbe^{*}_{2j-1}    \,.
\end{equation}
This means that  we can use the coefficients \eqref{CoefD} to redefine the elements of the fermionic basis and 
obtain purely  CFT objects : 
\begin{equation}
\bbe^*_{2m-1} = D_{2m-1}(a) \bbe^{\text{CFT}*}_{2m-1}\,, \quad\quad \bga^*_{2m-1} = D_{2m-1}(Q-a) \bga^{\text{CFT}*}_{2m-1} \,.
\end{equation}
For $\bbe^{\text{CFT}*}_{2m-1}$ and $\bga^{\text{CFT}*}_{2m-1}$ we have clear transformation rules under $\si_{1,2}$.
As in the non-super symmetric case for $\si_2$ 
\begin{equation}
\bbe^{\text{CFT}*}_{2m-1} \to \bga^{\text{CFT}*}_{2m-1} \,, \quad\quad \bga^{\text{CFT}*}_{2m-1} \to \bbe^{\text{CFT}*}_{2m-1}  \,.
\label{refl1}
\end{equation}
For $\si_1$ we must consider an additional term comming from the change in the passage from $D_{2m-1}(a)$ to $D_{2m-1}(Q-a)$ : 
\begin{equation}
D_{2m-1}(Q-a) = D_{2m-1}(-a) \left( \frac{a-(2m-1)b^{-1}}{a+(2m-1)b} \right) \,,
\end{equation}
which implies

\begin{align}
& \bbe_{2m-1}^{\text{CFT}*} \to \(\frac{a-(2m-1)b}{a+(2m-1)b^{-1}} \) \bga_{2m-1}^{\text{CFT}*} \label{refl2} \,, 
\\
& \bga_{2m-1}^{\text{CFT}*} \to \( \frac{a-(2m-1)b^{-1}}{a+(2m-1)b} \)  \bbe_{2m-1}^{\text{CFT}*} \nn \,.
\end{align}



The main conclusion drawn from Section \ref{subsec:Num}, is that the fermionic basis should be decomposable on the super Virasoro basis in the following way :
\begin{equation}
\bbe^{\text{CFT}*}_{I^+}\bga^{\text{CFT}*}_{I^-}V_a = C_{I^+,I^-}\( P^E_{I^+,I^-}(\{\bl_{-2k},\bs_{-r}\},\De,c)+d_a P^O_{I^+,I^-}(\{\bl_{-2k},\bs_{-r}\},\De,c)  \)V_a \,,
\label{decV}
\end{equation}
where  $C_{I^+,I^-}$ is the Cauchy determinant and $d_a$ is the function \eqref{dOdd} rewritten in the variables $a,b$ : 
\begin{equation}
d_a = \frac{1}{8}\sqrt{\left(9- \widehat{c} \right) \left(16\Delta_\alpha+1-\widehat{c} \right)} = \frac{1}{4}(b-b^{-1})(Q-2a) \,.
\end{equation}
The functions $P^E_{I^+,I^-}$ and $P^O_{I^+,I^-}$ ($E,O$ subscripts stand respectively for even and odd)
are polynomials in the modes of the super Virasoro algebra, depending rationally on the parameters $\De,c$. They are defined modulo the local integrals
$\bi_{2k-1}$  and satisfy the symmetry relations :
\begin{equation}
P^E_{I^+,I^-}  = P^E_{I^-,I^+} \,, \quad\quad P^O_{I^+,I^-}=-P^O_{I^-,I^+}             \,.
\end{equation}
The decomposition \eqref{decV}, as well as the transformation rules \eqref{refl1} and \eqref{refl2}, imply 
a relation of the type
\begin{align}
& \bbe^{\text{CFT}*}_{I^+}\bga^{\text{CFT}*}_{I^-}V_a = C_{I^+,I^-} \prod_{2j-1\in I^+} (a+(2j-1)b^{-1})  \prod_{2j-1\in I^-} (a+(2j-1)b)
\nn \\
& \times\( Q^E_{I^+,I^-}(\{a_{-2k},b_{-r}\},b_r\},a^2,Q^2)+g_a Q^O_{I^+,I^-}(\{a_{-2k},b_{-r}\},a^2,Q^2)  \)V_a,
\label{decH}
\end{align}
 with 
\begin{equation}
g_a = a (b-b^{-1}) \nn \,,
\end{equation}
and $Q^E_{I^+,I^-}$, $Q^O_{I^+,I^-}$ polynomials in the super Heisenberg algebra, depending rationally on $a^2$ and $Q^2$.
In the following we are going to verify this conjecture 
level by level.
\\
\subsubsection*{Level 2}
Let us start with the simplest case of level 2 :
\begin{equation}
\bbe^{\text{CFT}*}_{1}\bga^{\text{CFT}*}_{1}V_a = \Om_{1,1} = P^2 - \frac{1}{16} - \frac{\De_a}{8} \,. \label{reOm11}
\end{equation}
%
On this level only two operators $\bl_{-2}$ and $\bs_{-\frac{3}{2}}\bs_{-\frac{1}{2}}  $ are present. The calculation of one point functions 
on the cylinder was explained in Subsection \ref{subsec:Descendants} and gave in this case \eqref{1PFLvl2}:  
\begin{equation}
\braket{\bl_{-2}V_a} = \de_P - \frac{c}{24} - \frac{\De_a}{12} \, ,
\quad\quad 
\braket{\bs_{-\frac{3}{2}}\bs_{-\frac{1}{2}}} = \frac{\De_a}{12} \,,
\end{equation}
Hence it is not difficult to compare with \eqref{reOm11} to obtain : 
\begin{equation}
\bbe^{\text{CFT}*}_{1}\bga^{\text{CFT}*}_{1}V_a = \(  \bl_{-2} - \frac{1}{2} \bs_{-\frac{3}{2}}\bs_{-\frac{1}{2}}  \)V_a \,.
\label{lev2}
\end{equation}
Using \eqref{ULvl2}, one can rewrite the combination \eqref{lev2} as : 
\begin{equation}
\bbe^{\text{CFT}*}_{1}\bga^{\text{CFT}*}_{1}V_a = \frac{1}{4}(a+b)(a+b^{-1}) \left( a_{-1}^2 +2 b_{-\frac{3}{2}}b_{-\frac{1}{2}} \right)V_a \,.
\end{equation}
This neat factorisation of the term $(a+b)(a+b^{-1}) $ is a check of our conjecture, and the above shows that :
$Q^E_{\{1,1\}} =\frac{1}{4} \left( (a_{-1})^2 +2 b_{-\frac{3}{2}}b_{-\frac{1}{2}} \right) $.
\\
\\
The main difference with the usual Liouville case,
is that at higher levels, we do not know a priori the decompositions of the type \eqref{decV} (recall the discussion at the end of the Section \ref{subsec:Num}).
To overcome this difficulty, we shall proceed as in \cite{NeSm} and obtain  the decompostion by solving the reflection constraints implied by \eqref{decH}.  
Let us briefly recall the main steps.

Consider that at an (even) level $k=|I^+|+|I^-|$ we have a basis of super Virasoro generators $\{\bv^{(k)}_1,...,\bv^{(k)}_d \}$ (by convention we consider that
$\bv^{(k)}_1 = \bl_{-2}^{\frac{k}{2}}$ ) that are 
related to the super Heisenberg basis $\{\bh_1^{(k)},...,\bh_d^{(k)}\}$ modulo the action of integrals of motions by : 
\begin{equation}
\bv^{(k)}_i = \sum_{k=1}^d U_{i,j}^{(k)}(a) \bh_j^{(k)}      \nn \,,
\end{equation}
with $U^{(k)}(a)$ the passage matrix, whose determinant is factorisable  : 
\begin{equation}
 \mathrm{det}(U^{(k)}(a)) = C^{(k)} N^{(k)}(a,b) \frac{D_V^{(k)}(\De,c)}{D_H^{(k)}(a^2,Q^2)} \,,
\end{equation}
where $N^{(k)}(a,b)$ is the null vector contribution. We look for $P^E_{I^+,I^-},\,P^E_{I^+,I^-} $ in the form :  
\begin{align}
P^E_{I^+,I^-} & = \bv_1+\frac{1}{D^{(k)}_V(\De,c)} \sum_{i=2}^d X_{I^+,I^-,i}(\De,c)\bv_i  \,, \nn\\
P^O_{I^+,I^-} & = \frac{1}{D^{(k)}_V(\De,c)} \sum_{i=2}^d Y_{I^+,I^-,i}(\De,c)\bv_i \,,   \nn
\end{align}
where $X_{I^+,I^-,i}(\De,c),\,Y_{I^+,I^-,i}(\De,c)$ are polynomials of some degree $D$ to be determined. Also introduce the polynomials :
\begin{align}
  T^+_{I^+I^-}(a) & = \frac{1}{2}\left( \prod_{j\in I^+}(a+jb^{-1})\prod_{j\in I^-}(a+jb)
+ \prod_{j\in I^+}(a+jb)\prod_{j\in I^-}(a+jb^{-1})      \right)  
\,,\nn \\
  T^-_{I^+I^-}(a) & = \frac{1}{2(b-b^{-1})}\left( \prod_{j\in I^+}(a+jb^{-1})\prod_{j\in I^-}(a+jb)
- \prod_{j\in I^+}(a+jb)\prod_{j\in I^-}(a+jb^{-1})      \right)   \nn \,.
\end{align}
Then \eqref{decH} gives strong conditions on the structure of $X_{I^+,I^-,i}(\De,c),\,Y_{I^+,I^-,i}(\De,c)$ (see \cite{NeSm} for details).
For any $1\leq j\leq d$ we must have  
\begin{align}
&  D^{(k)}_V(\De(-a),c) D^{(k)}_H(a^2,Q^2) \label{Cnst1} \\
& \times \{ T^+_{I^+I^-}(-a) \( D^{(k)}_V(\De,c) U_{1,j}^{(k)}(a)+ \sum_{i=2}^d X_{I^+,I^-,i}U_{i,j}^{(k)}(a)      \) \nn\\
& -(Q^2-4)(Q-2a) T^-_{I^+I^-}(-a) \sum_{i=2}^d Y_{I^+,I^-,i} U_{i,j}^{(k)}(a) \} \quad  \text{is even in $a$,}  \nn
\end{align}
and 
\begin{align}
&  D^{(k)}_V(\De(-a),c) D^{(k)}_H(a^2,Q^2) \label{Cnst2} \\
& \times \{ -T^-_{I^+I^-}(-a) \( D^{(k)}_V(\De,c) U_{1,j}^{(k)}(a)+ \sum_{i=2}^d X_{I^+,I^-,i}U_{i,j}^{(k)}(a)      \) \nn\\
& +(Q^2-4)(Q-2a) T^+_{I^+I^-}(-a) \sum_{i=2}^d Y_{I^+,I^-,i} U_{i,j}^{(k)}(a) \} \quad  \text{is odd in $a$,}  \nn
\end{align}
Taking the degree $D$ appropriately large, we obtain enough linear equations on the coefficients of $X_{I^+,I^-,i}(\De,c),\,Y_{I^+,I^-,i}(\De,c)$.
Now we demonstrate how this procedure works at higher levels.
\\
\subsubsection*{Level 4}
Consider the set up described in \ref{subsubsec:Lvl4}. Recall that at this level there are 5  operators in total (modulo the action of $\bi_1$),
that are : 
\begin{equation}
\bl_{-2}^2\,, \quad\quad \bl_{-4}\,, \quad\quad
\bs_{-\frac{7}{2}}\bs_{-\frac{1}{2}}\,,  \quad\quad \bs_{-\frac{5}{2}}\bs_{-\frac{3}{2}}\,, \quad\quad
\bl_{-2}\bs_{-\frac{3}{2}}\bs_{-\frac{1}{2}} \,. \nn
\end{equation}
We solve the constraints \eqref{Cnst1} and \eqref{Cnst2} with the use of \eqref{U4} and \eqref{detU4}, and obtain the following expressions : 
\begin{align}
P^E_{\{1,3\}} & = \bl_{-2}^2 + \left( \frac{-45+4c}{18} -\frac{\De_a}{3} \right) \bl_{-4} + \left( \frac{45-4c}{36} +\frac{\De_a}{6} \right)
\bs_{-\frac{7}{2}}\bs_{-\frac{1}{2}} + \nn \\ & \frac{1}{4}\bs_{-\frac{5}{2}}\bs_{-\frac{3}{2}} -\frac{1}{2}\bl_{-2}\bs_{-\frac{3}{2}}\bs_{-\frac{1}{2}} \,, \nn
\\
P^O_{\{1,3\}} & = \frac{1}{3}\bl_{-4}-\frac{1}{6} \bs_{-\frac{7}{2}}\bs_{-\frac{1}{2}} \,, \nn
\end{align}
as well as the mirror polynomials $P^E_{\{3,1\}},P^O_{\{3,1\}}$.
One can now compute the one point function  of $  \left(P^E_{\{1,3\}\atop{\{3,1\}}}\mp d_a P^O_{\{1,3\}\atop{\{3,1\}}}\right) V_a  $, all the individual
contributions of descendants at level 4 are given in \eqref{1PFLvl4}. One recovers exactly the value of $\Om_{{1,3}\atop{3,1}}  $  obtained in  \eqref{Om13} by interpolation.
Summarising : 
\begin{equation}
\boldsymbol{\beta}_{1\atop{3}}^{\text{CFT}*} \boldsymbol{\gamma}_{3\atop{1}}^{\text{CFT}*} V_a  =  \Om_{{1,3}\atop{3,1}}  
= \frac{1}{2} \(P^E_{\{1,3\}\atop{\{3,1\}}}(\{\bl_{-2k},\bs_{-r}\},\De,c) \mp d_a P^O_{\{1,3\}\atop{\{3,1\}}} (\{\bl_{-2k},\bs_{-r}\},\De,c)\)V_a  \,.
\end{equation}
This is an independent argument in favor of \eqref{Om13}.

\subsubsection*{Level 6}
We proceed through the same analysis. Recall that we have (modulo the action of $\bi_1$ and $\bi_3$ )
10 Virasoro operators, that we took to be
\begin{align}
&\bl_{-2}^3\,, \quad \bl_{-6}\,, \quad \bl_{-3}^2\,, \quad  \bs_{-\frac{7}{2}} \bs_{-\frac{5}{2}}\,,  
\quad  \bs_{-\frac{9}{2}} \bs_{-\frac{3}{2}}\,, \quad  \bs_{-\frac{11}{2}} \bs_{-\frac{1}{2}} \,,\nn\\
& \bl_{-2}^2\bs_{-\frac{3}{2}}\bs_{-\frac{1}{2}}\,, \quad \bl_{-2}\bs_{-\frac{7}{2}}\bs_{-\frac{1}{2}}\,, 
\quad \bl_{-2}\bs_{-\frac{5}{2}}\bs_{-\frac{3}{2}}\,, \quad \bl_{-3}\bs_{-\frac{5}{2}}\bs_{-\frac{1}{2}} \nn \,.
\end{align}
Using the explicit value of $U^{(6)}$ and the factors \eqref{detU6}, the reflection constraints 
bring the following results : 
\begin{align}
& P^E_{\{3,3\}}  =                  \bl_{-2}^3 + 
                      \frac{1}{480} \left(572 \Delta _{a }^2+1976 \Delta _{a }-80 c^2-96 c \Delta _{a }+2076
   c-18381\right)\bl_{-6}  + \nn\\&
\frac{1}{96} \left(12 \Delta _{a }^2+228 \Delta _{a }-16 c \Delta _{a }-12
   c-27\right)\bl_{-3}^2 + \nn\\&
\frac{1}{192} \left(-28 \Delta _{a }^2+192 \Delta _{a }-16 c \Delta _{a }-20
   c+117\right)\bs_{-\frac{7}{2}} \bs_{-\frac{5}{2}}  + \nn\\&
\frac{1}{64} \left(-4 \Delta _{a }^2+92 \Delta _{a }-8 c \Delta _{a }-8
   c+105\right)\bs_{-\frac{9}{2}} \bs_{-\frac{3}{2}} + \nn\\&
\frac{1}{960} \left(28 \Delta _{a }^2+404 \Delta _{a }+56 c \Delta _{a }-136
   c+6021\right)\bs_{-\frac{11}{2}} \bs_{-\frac{1}{2}} +   \nn\\&
-\frac{1}{2} \bl_{-2}^2\bs_{-\frac{3}{2}}\bs_{-\frac{1}{2}} + \frac{1}{12} \left(4 \Delta _{a }-2 c+27\right)\bl_{-2}\bs_{-\frac{7}{2}}\bs_{-\frac{1}{2}} +  \nn\\&
  \frac{1}{16} \left(9-2 \Delta _{a   }\right)        \bl_{-2}\bs_{-\frac{5}{2}}\bs_{-\frac{3}{2}}  + 
 \frac{1}{192} \left(4 \Delta _{a }^2-68 \Delta _{a }+8 c \Delta _{a }-8 c+291\right) \bl_{-3}\bs_{-\frac{5}{2}}\bs_{-\frac{1}{2}} \nn \,,
\end{align}
as well as
\begin{align}
& P^E_{\{1,5\}}  =                  \bl_{-2}^3 +   \frac{1}{90} \left(79 \Delta _{a }^2+1004 \Delta _{a }-12 c^2-98 c \Delta _{a }+322
   c-2855\right)\bl_{-6}  + \nn\\&
     \frac{1}{12} \left(2 \Delta _{a }^2+12 \Delta _{a }-2 c \Delta _{a }-c-2\right) \bl_{-3}^2 + %
     \frac{1}{36}  \left(\Delta _{a }^2+7 \Delta _{a }-2 c \Delta _{a}-3 c+18\right)\bs_{-\frac{7}{2}} \bs_{-\frac{5}{2}}  + \nn\\&
     \frac{1}{8} \left(2 \Delta _{a }-c+11\right)\bs_{-\frac{9}{2}} \bs_{-\frac{3}{2}} + 
     \frac{1}{180} \left(-19 \Delta _{a }^2-804 \Delta _{a }+2 c^2+88 c \Delta _{a }-72
   c+1315\right)\bs_{-\frac{11}{2}} \bs_{-\frac{1}{2}} +   \nn\\&
      -\frac{1}{2}\bl_{-2}^2\bs_{-\frac{3}{2}}\bs_{-\frac{1}{2}} + 
      \frac{1}{9} \left(2 \Delta _{a }-c+14\right)\bl_{-2}\bs_{-\frac{7}{2}}\bs_{-\frac{1}{2}} +  \nn\\&
      \frac{1}{2}\bl_{-2}\bs_{-\frac{5}{2}}\bs_{-\frac{3}{2}}  + 
      \frac{1}{36} \left(-\Delta _{a}^2+\Delta _{a}+2 c \Delta _{a }-c+38\right) \bl_{-3}\bs_{-\frac{5}{2}}\bs_{-\frac{1}{2}} \nn \,.
\end{align}
Finally we obtain
\begin{align}
& P^O_{\{1,5\}}  =            
                     \frac{1}{30} \left(-136 \Delta _{a }-12 c+335\right) \bl_{-6}  + 
                      -\Delta _{a }\bl_{-3}^2 + 
                      \frac{1}{12} \left(-4 \Delta _{a}-3\right)\bs_{-\frac{7}{2}} \bs_{-\frac{5}{2}}  + \nn\\&
                       -\frac{3}{4}\bs_{-\frac{9}{2}} \bs_{-\frac{3}{2}} + 
                      \frac{1}{60} \left(76 \Delta _{a }-8 c-115\right)\bs_{-\frac{11}{2}} \bs_{-\frac{1}{2}} +  
                       -\frac{2}{3}\bl_{-2}\bs_{-\frac{7}{2}}\bs_{-\frac{1}{2}} +  
                      \frac{1}{12} \left(4 \Delta_{a }-5\right)\bl_{-3}\bs_{-\frac{5}{2}}\bs_{-\frac{1}{2}} \nn \,.
\end{align}
We also find the same expressions of  the polynomials $P^E_{\{5,1\}} $ and $ P^O_{\{5,1\}}$ (up to a relevant minus sign for $ P^O_{\{5,1\}}$).
Then one can proceed and calculate the relevant one point functions of descendants on the cylinder (see Subsection \ref{subsec:Descendants}). We summarize here the results :  
\begin{align}
\braket{\bl_{-6} V_a}  & = -\frac{\Delta _{a }}{6048}\,, \quad \quad  \braket{\bl_{-3}^2V_a}  = \frac{72 \Delta _{a }+31 c-504 \delta _P}{30240}  \nn \,, \\
 \nn  \\ 
\braket{\bs_{-\frac{7}{2}} \bs_{-\frac{5}{2}} V_a}  & = \frac{604 \Delta _{a }+457 c-3528 \delta _P}{483840} \,,
\quad\quad
\braket{\bs_{-\frac{9}{2}} \bs_{-\frac{3}{2}}V_a}   = \frac{-1371 \Delta _{a }-457 c+3528 \delta _P}{1451520}
\nn \,, \\
\braket{\bs_{-\frac{11}{2}} \bs_{-\frac{1}{2}} V_a}  & = \frac{31 \Delta _{a }}{96768} \,, 
\quad\quad
\braket{\bl_{-2}\bs_{-\frac{7}{2}}\bs_{-\frac{1}{2}} V_a}  = 
\frac{294 \Delta _{a }^2+1252 \Delta _{a }+147 c \Delta _{a }-3528 \Delta _{a } \delta _P}{483840}
\nn \,, \\
\braket{\bl_{-3}\bs_{-\frac{5}{2}}\bs_{-\frac{1}{2}} V_a} & = \frac{17 \Delta _{a }}{60480}\nn \,, 
\end{align}
as well as the most complex results : 
\begin{align}
 & \braket{\bl_{-2}^3 V_a}  = \frac{1}{483840}    \Big(   -280 \Delta _{a}^3-2352 \Delta _{a }^2-3968 \Delta _{a }-35 c^3-210 c^2 \Delta _{a }+2520 c^2 
 \nn  \\ &
 \delta _P-462 c^2-420 c \Delta _{a }^2-2100 c \Delta _{\alpha }+ 
    10080 c \Delta _{a } \delta _P-60480 c \delta _P^2+21168
   c \delta _P-1504 c+ 
   \nn \\ &
   10080 \Delta _{a }^2 \delta _P-120960 \Delta _{a } \delta _P^2+48384 \Delta _{a} \delta_P+483840 \delta _P^3-241920 \delta _P^2+32256 \delta _P \Big) 
   \nn \\
& \braket{\bl_{-2}^2\bs_{-\frac{3}{2}}\bs_{-\frac{1}{2}} V_a} =
\frac{\Delta _{a }}{241920}\Big(140 \Delta _{a }^2+ 
672 \Delta _{a}+35 c^2+140 c \Delta _{a }-1680 c \delta _P+
\nn\\ &
294 c-3360 \Delta _{a } \delta _P 
    +20160 \delta _P^2-6720 \delta _P+544  \Big)     \,,  \nn
\nn\\ 
&\braket{\bl_{-2}\bs_{-\frac{5}{2}}\bs_{-\frac{3}{2}}V_a} =
\frac{1}{483840} \Big( -714 \Delta _{a }^2-3588 \Delta _{a}-119 c^2-
\nn\\&
595 c \Delta _{a}+4536 c \delta _P-1196 c+11928 \Delta
   _{a } \delta _P-40320 \delta _P^2+18144 \delta _P \Big) \,. \nn
\end{align}
Using these values for the one point functions, we recover exactly the expressions \eqref{Om33} and \eqref{Om15}. That is we check that : 
\begin{align}
 \boldsymbol{\beta}_{1\atop{5}}^{\text{CFT}*} \boldsymbol{\gamma}_{5\atop{1}}^{\text{CFT}*} V_a  &=  \Om_{{1,5}\atop{5,1}}  = 
\frac{1}{3} \(P^E_{{\{1,5\}}\atop{\{5,1\}}}(\{\bl_{-2k},\bs_{-r}\},\De,c)\mp d_a P^O_{{\{1,5\}}\atop{\{5,1\}}}(\{\bl_{-2k},\bs_{-r}\},\De,c)\)V_a
\, \\
\boldsymbol{\beta}_3^{\text{CFT}*} \boldsymbol{\gamma}_3^{\text{CFT}*} V_a  &=  \Om_{1,3}  = \frac{1}{3} P^E_{\{3,3\}}(\{\bl_{-2k},\bs_{-r}\},\De,c)V_a   \,.
\end{align}
This strongly confirms the results obtained by interpolation.

\section{Conclusion}
The achievement of this work is the computation of the one point functions of fermionic operators in the ssG model.
They are constructed out of a single function $\Om$, defined by a set of scaling equations, and which origin is traced to the computation of vacuum expectation values 
of lattice operators on the underlying 19-vertex model. 
On one hand, the analysis of the scaling equations in the conformal regime allowed to compute the one point functions of specific fermionic operators in the UV limit,
and to establish the correspondence between the usual Virasoro description of CFT and the fermionic part of the fermion-current description.
On the other hand, these results have been checked by an alternative method that relies on the reflection symmetry of the ssG model. 
We emphasize again that both techniques completely differ in their nature and are both based on conjectures. 
The matching of the results from both sides is a very strong assertion for both of them.

{Concerning the primary fields notice that 
we have obtained 
the most important for applications 
quanitity. Indeed, we argued   that  the simplest non-chiral fermionic descendant
provides the ratio of one point functions of the operators
$W_{\al+\frac{2\beta^2}{1-\beta^2}}$ 
and $V_\al$. The former operator is exactly the most
relevant contribution  occurring  in the OPE of the latter one with the
perturbing
operator $W_{\frac{2\beta^2}{1-\beta^2}}$. In other words the ratio of one-point functions
in question provides the most important contribution to the conformal
perturbation theory.}
\\
\\
We need to consider the entire space of local operators adding those created by the KM currents.
The one-point functions of the latter include the function $\omega(\theta,\theta')$.
Recall the equation \eqref{defOmega}.
Using this equation and known $\Omega(\theta,\theta')$ one can, in principle, reconstruct $\omega(\theta,\theta')$.
The result is not unique, one has to find a way of fixing the quasi-constants (anti-periodic  with period $\pi i$ 
functions of $\theta,\theta'$. Following this numerically it is hard to achieve a good precision which makes it
difficult to put forward a conjecture based on the interpolation. This is a technical difficulty which
we hope to overcome in future.


\end{document}